\documentclass[usenatbib]{mn2e}
\usepackage{graphicx}
\usepackage{amsbsy}
\usepackage{xspace}
\usepackage{subfigure}
\usepackage{natbib}
\usepackage{color}

\usepackage[colorinlistoftodos]{todonotes}
\usepackage[normalem]{ulem}
\usepackage{booktabs,caption}
\usepackage[flushleft]{threeparttable}
\def\lesssim{\mathrel{\hbox{\rlap{\hbox{\lower5pt\hbox{$\sim$}}}\hbox{$<$}}}}
\def\gtrsim{\mathrel{\hbox{\rlap{\hbox{\lower5pt\hbox{$\sim$}}}\hbox{$>$}}}}
\def\kms{{\rm km\,s$^{-1}$}\xspace}

\def\apjl{ApJ}%
\def\apjs{ApJS}%
\def\aap{A\&A}%
\title[iPTF $g+I$ band Survey]{iPTF Survey for Cool Transients}
\author[Adams et al.]
  {\parbox{18cm}{S.~M.~Adams$^{1}$, N.~Blagorodnova$^{1}$, M.~M.~Kasliwal$^{1}$, R.~Amanullah$^{2}$, T.~Barlow$^{1}$, B.~Bue$^{3}$, M.~Bulla$^{2}$, Y.~Cao$^{4}$, S.~B.~Cenko$^{5,6}$, D.~O.~Cook$^{1}$, R.~Ferretti$^{2}$, 
O.~D.~Fox$^{7}$,
 C.~Fremling$^{8}$, 
S.~Gezari$^{6,9}$,
A.~Goobar$^{2}$, A.~Y.~Q.~Ho$^{1}$, T.~Hung$^{8}$, E.~Karamehmetoglu$^{8}$, S.~R.~Kulkarni$^{1}$, T.~Kupfer$^{1}$, R.~R.~Laher$^{10}$, F.~J.~Masci$^{10}$, A.~A.~Miller$^{11,12}$, J.~D.~Neill$^{1}$, P.~E.~Nugent$^{13,14}$, J.~Sollerman$^{8}$, F.~Taddia$^{8}$, \& R.~Walters$^{1}$}
  \\
  \\
  $^{1}$ Cahill Center for Astrophysics, California Institute of Technology, Pasadena, CA 91125, USA\\
  $^{2}$ The Oskar Klein Centre, Department of Physics, Stockholm University, AlbaNova, SE-106 91 Stockholm, Sweden \\
  $^{3}$ Jet Propulsion Laboratory, California Institute of Technology, Pasadena, CA 91109, USA \\
  $^{4}$ eScience Institute and Astronomy Department, University of Washington, Seattle, WA 98195, USA \\
  $^{5}$ Astrophysics Science Division, NASA Goddard Space Flight Center, Code 661, Greenbelt, MD 20771, USA \\
  $^{6}$ Joint Space-Science Institute, University of Maryland, College Park, MD 20742, USA \\
  $^{7}$ Space Telescope Science Institute, 3700 San Martin
Dr, Baltimore, MD 21218 \\
  $^{8}$ The Oskar Klein Centre, Department of Astronomy, AlbaNova, SE-106 91 Stockholm, Sweden \\ 
  $^{9}$ Department of Astronomy, University of Maryland, College Park, MD 20742, USA \\
  $^{10}$ Infrared Processing and Analysis Center, California Institute of Technology, Pasadena, CA 91125, USA \\
  $^{11}$ Center for Interdisciplinary Exploration and Research in Astrophysics (CIERA) and Department of Physics and Astronomy, \\ ~~~~~~~~Northwestern University, 2145 Sheridan Road, Evanston, IL 60208, USA \\
  $^{12}$ The Adler Planetarium, Chicago, IL 60605, USA \\
  $^{13}$ Department of Astronomy, University of California, Berkeley, CA 94720-3411, USA \\
  $^{14}$ Lawrence Berkeley National Laboratory, 1 Cyclotron Road, MS 50B-4206, Berkeley, CA 94720, USA \\
  E-mail: sma@astro.caltech.edu}

\begin{document}
\voffset -1.5cm
\maketitle

\begin{abstract}
We performed a wide-area (2000 deg$^{2}$) $g$ and $I$ band experiment as part of a two month extension to the Intermediate Palomar Transient Factory.  We discovered 36 extragalactic transients including iPTF17lf, a highly reddened local SN Ia, iPTF17bkj, a new member of the rare class of transitional Ibn/IIn supernovae, and iPTF17be, a candidate luminous blue variable outburst.  We do not detect any luminous red novae and place an upper limit on their rate.
We show that adding a slow-cadence $I$ band component to upcoming surveys such as the Zwicky Transient Facility will improve the photometric selection of cool and dusty transients. 
\\
\\
\end{abstract}

\section{Introduction}
Transient astronomy is benefiting enormously from the deployment of increasingly capable wide-field cameras in time-domain surveys such as ASAS-SN \citep{Shappee14}, ATLAS \citep{Tonry11}, MASTER \citep{Gorbovskoy13}, the Palomar Transient Factory \citep[PTF;][]{Law09}, and Pan-STARRS \citep{Chambers16}.  However, with the upcoming Zwicky Transient Facility (ZTF; \citealt{Bellm14}) and, later, the Large Synoptic Survey Telescope \citep{LSSTsciencebook}, the torrent of transient discoveries will vastly exceed spectroscopic follow-up capabilities.  
While some interesting transients can be prioritized for spectroscopic follow-up on the basis of rapidly evolving light curves from high-cadence experiments, for many science cases 
it will only be practical to select targets based on photometric colors.

Despite the impending need for the next generation of transient surveys to utilize multiple filters concurrently, such an approach has generally been limited to transient surveys with relatively small footprints (e.g., the Pan-STARRS Medium Deep Survey -- \citealt{Chambers16}; the DES SN survey -- \citealt{Kessler15}; SNLS -- \citealt{Howell05,Astier06}. SDSS II Supernova Survey -- \citealt{Frieman08}).
Recently, \citet{Miller17} described the first quasi-contemporaneous multi-band ($g$ and $R$) survey to be undertaken with the Intermediate Palomar Transient Factory \citep[iPTF;][]{Kulkarni13}.

Extragalactic transient surveys have generally avoided $I$ band because quickly evolving transients are hot (at least initially) and the sky background is higher in $I$ band than in bluer optical filters.
Still, adding $I$ band imaging as an additional filter to a transient survey may provide a number of benefits.

An $I$ band filter can increase the sensitivity to and improve the selection of cool or dusty events such as luminous red novae (LRNe), intermediate luminosity red transients (ILRTs), and obscured SNe.  This is illustrated by Fig. 1, which shows the color evolution of representatives of various classes of transients.  While most SNe have modest ($g-I\lesssim 0.5$ mag) colors at early times, a significant fraction ($\sim$19\%; \citealt{Mattila12}) of local SNe are so heavily reddened that they are missed by standard optical surveys.

LRNe and ILRTs are loosely defined observational classes of transients with peak magnitudes between novae and SNe Ia ($-10 \lesssim M \lesssim -16$), red optical colors, and low inferred ejecta velocities ($\lesssim1000~\mathrm{km}\>\mathrm{s}^{-1}$).  ILRTs (also referred to as SN~2008S-like transients) arise from obscured massive stars \citep{Thompson09}.  It has been suggests that ILRTs may be the result of electron-capture SNe \citep{Botticella09}, mass transfer episodes \citep{Kashi10}, or some other non-terminal outburst \citep{Berger09,Smith09}.  The most recent evidence from the late-time evolution of the best-studied events suggests that they are supernovae \citep{Adams16}. 
LRNe leave infrared-bright remnants and are of particular interest because they have been shown to be associated with stellar mergers or the onset of common envelope events \citep{Tylenda11,Ivanova13}.  Although they are common \citep{Kochanek14d}, their low luminosities and red colors ($g-I\gtrsim 1$ mag) have limited the number of identified candidates to a handful of events in the Milky Way \citep{Martini99,Soker03,Tylenda11,Tylenda13} and nearby ($\lesssim 7$ Mpc) galaxies \citep{Rich89,Kulkarni07,Williams15,Smith16,Blagorodnova17}.  
More events are needed to characterize the range of LRN and ILRT properties.

Furthermore, the addition of $I$ band survey data is important for SN Ia cosmology.  SN Ia distances are best determined with at least two colors \citep{Riess96,Jha07}.  
Employing $I$ band for one of these colors has the added benefit that the distinctive 2nd peak seen at $I$ band and redder filters (see Fig. 1) for a majority of (low redshift) SNe Ia facilitates photometric typing \citep{Poznanski02}.    

In this paper we describe a 2 month long experiment undertaken with the Palomar 48'' Telescope observing a wide area at low cadence in $g$ and $I$ bands.  
We present the 36 transients discovered, their spectroscopic classifications, rate estimates, and a description of three interesting transients discovered: a heavily reddened SN Ia, a transitional SN Ibn/IIn, and an LBV outburst.
All magnitudes in this paper are in the AB system.

\begin{figure}
\includegraphics[width=1.05\columnwidth]{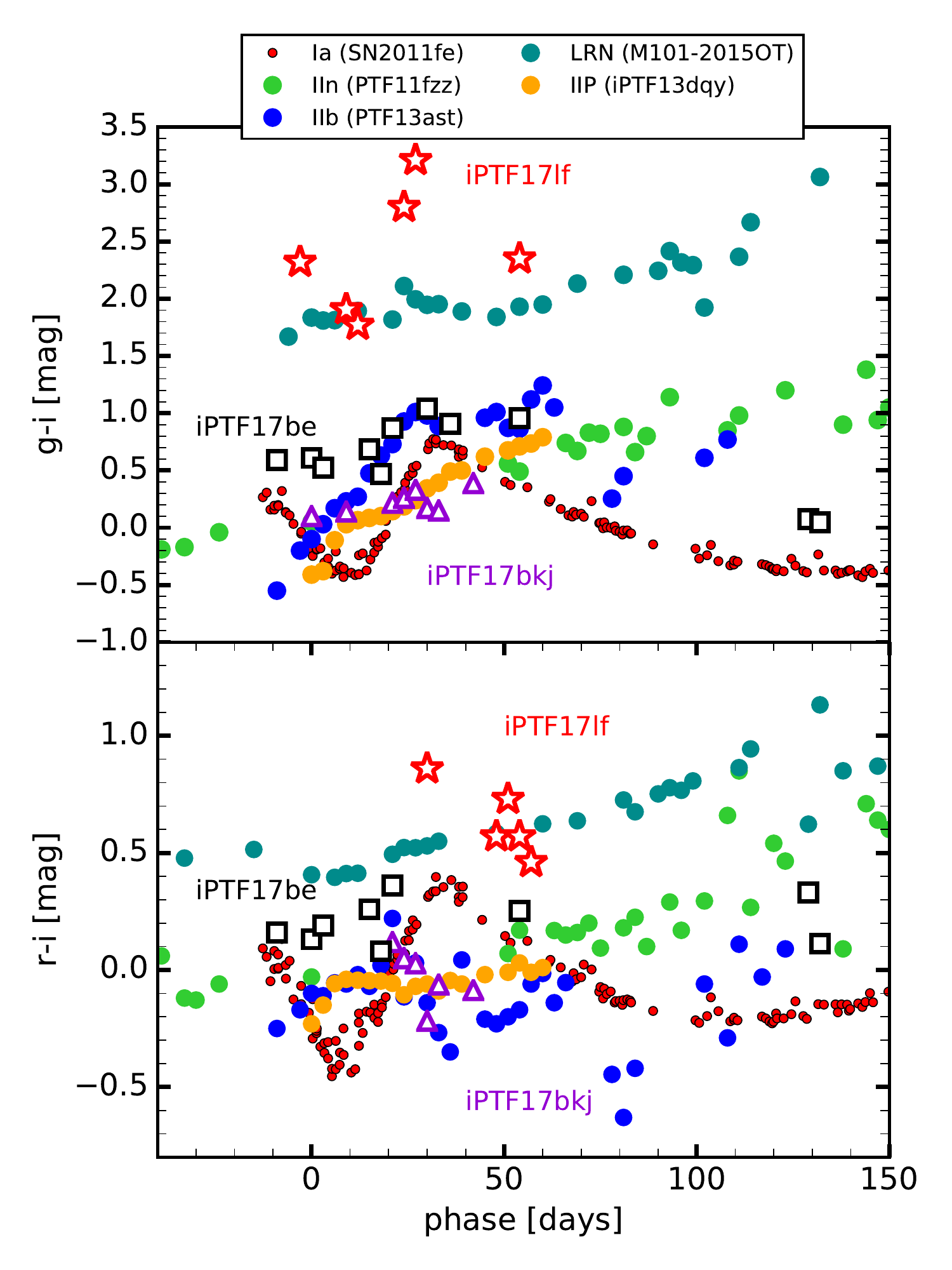}
\caption{Color evolution with phase for representatives of various classes of transients: SN IIP iPTF13dqy \citep{Yaron17}), SN Ia SN~2011fe \citep{Munari2013NewA}, SN IIn PTF11fzz \citep{Ofek2013Natur}, SN IIb iPTF13ast \citep{Gal-Yam2014Natur}, and LRN M101-OT \citep{Blagorodnova17}.  Including $I$ band in a time-domain survey can improve the photometric classification and selection of both SN Ia (by the late-time $I$ band bump) and very cool and/or dusty transients such as luminous red novae (by their extremely red color).  We also show three transients found in this survey: iPTF17lf (a highly reddened SN Ia; shown by the red stars), iPTF17be (a likely LBV outburst; black squares), and iPTF17bkj (a transitional SN Ibn/IIn; purple triangles).   \label{fig:gmini}}
\end{figure}

\section{Survey Design}
This experiment was undertaken as part of the extension to iPTF, which succeeded PTF \citep{Law09,Rau09}.  iPTF utilizes the Palomar 48" Telescope (P48) with the CFH12K camera \citep{Rahmer08,Law10} comprised of 11 working CCDs, each with 2048 $\times$ 4096 pixels, $1\farcs 01$ pixel$^{-1}$, for a 7.3 deg$^{2}$ field of view.  
The experiment was allocated 38 nights (of which observing conditions permitted data collection on 17 nights) during dark and gray time between 2017 January 7 and February 25, with the remainder of the nights used to complete the iPTF Census of the Local Universe H$\alpha$ survey \citep{Cook17} and follow up target of opportunity triggers. 

\begin{figure}
\begin{center}
\includegraphics[width=\columnwidth]{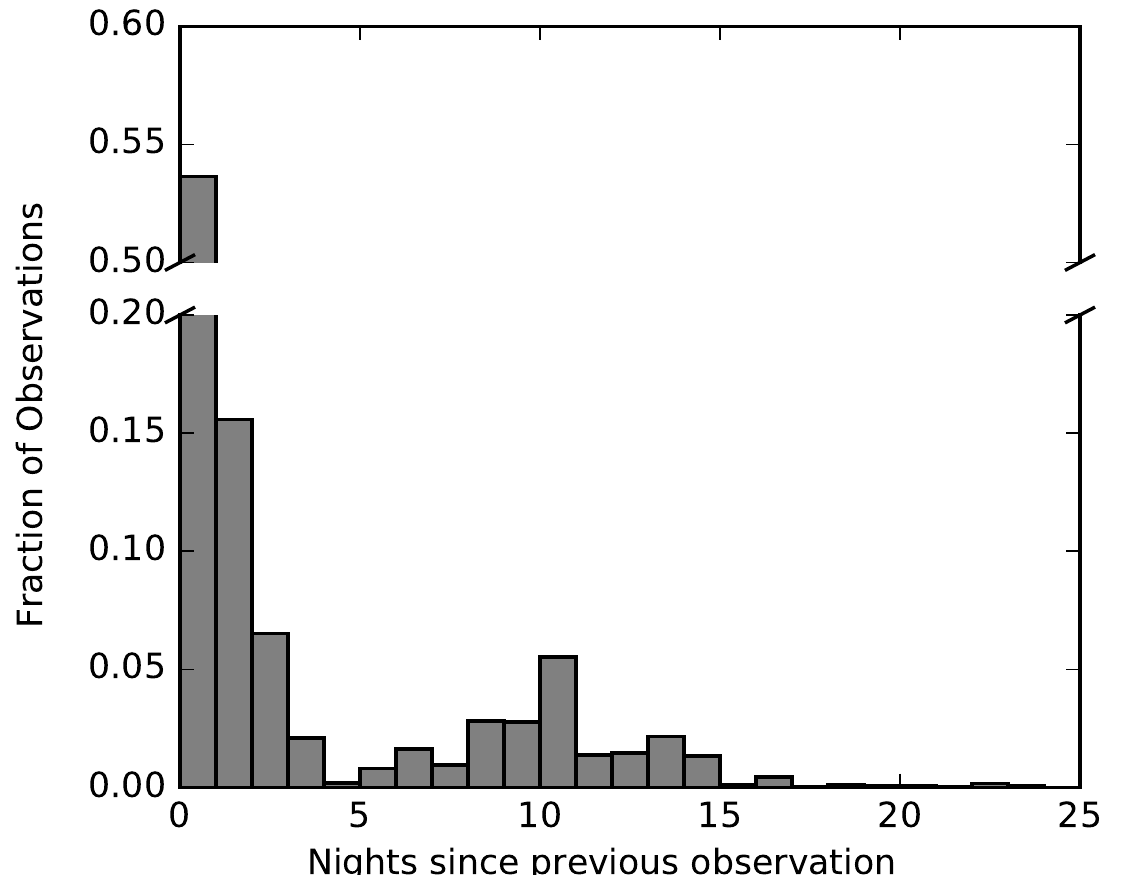}
\end{center}
\caption{Histogram showing the cadence distribution of the observations (in either filter).  The $\gtrsim50\%$ of observations with sub-night cadence represents the nightly $g$ and $I$ band observations of each targeted field.  During a full night devoted exclusively to the $g+I$ survey (and unhampered by weather) $\sim$50\% of the fields were observed.  
The slower tail in the cadence is the result of weather and other programs taking precedence according to lunation and ToO triggers. \label{fig:cadence}}
\end{figure}

\begin{figure}
\begin{center}
\includegraphics[width=\columnwidth]{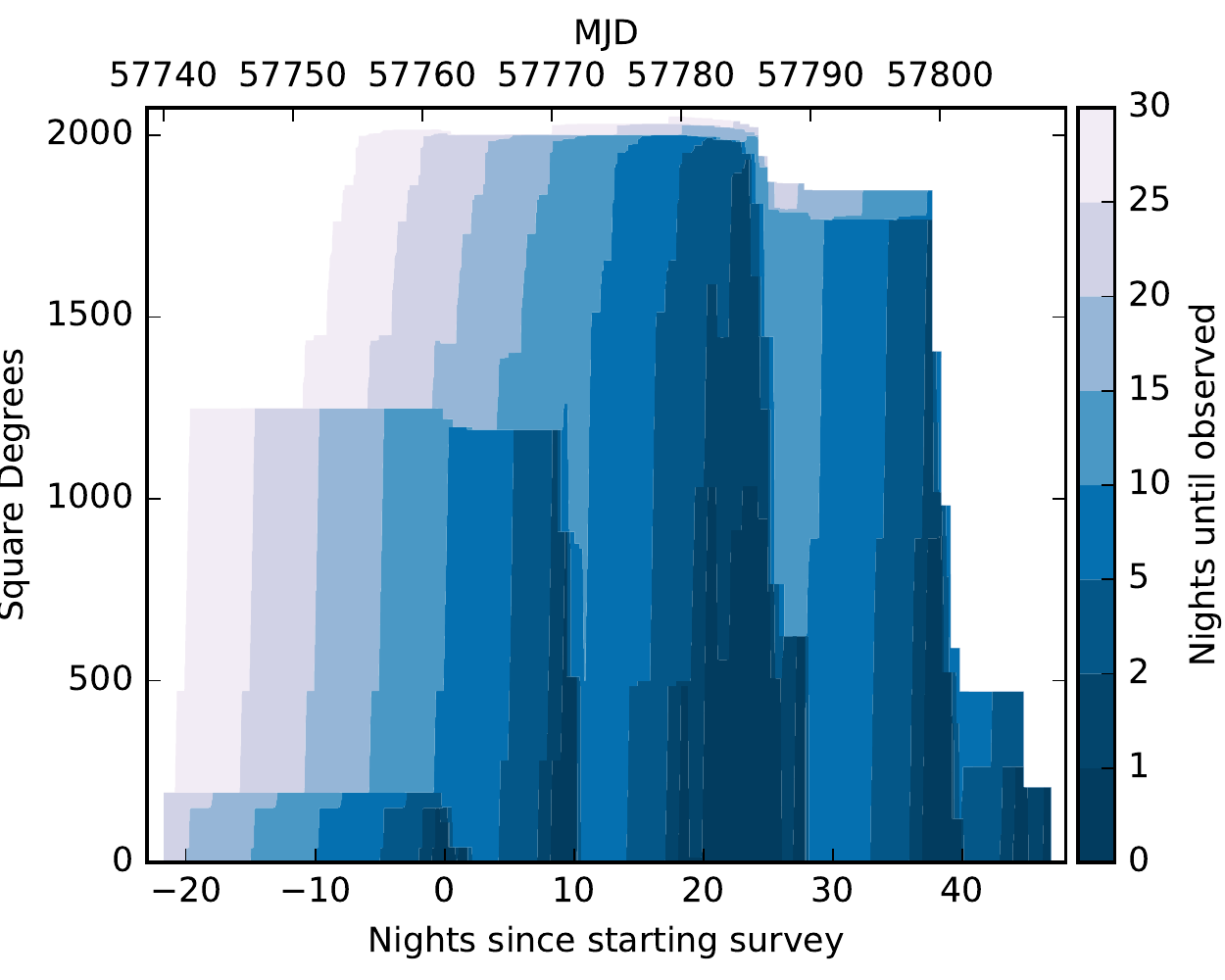}
\end{center}
\caption{Areal coverage throughout the survey rank-ordered by the time until survey fields are next observed (in both $g$ and $I$ bands within the same night).  The effective survey coverage (area$\times$time) for a transient with a given duration can be found by integrating the area under the given ``Nights until observed'' color contour, e.g., 327, 2170, and 3731 square-degree months for durations of 1, 10, and 30 days, respectively (by this metric coverage begins prior to the start of the survey since earlier transients are detected as long as they are still brighter than the survey limiting magnitude when their position is first covered by the survey).\label{fig:controltime}}
\end{figure}

The experiment monitored $\sim$2000 square degrees with fields chosen to avoid the Galactic and ecliptic planes and virtually no bias toward nearby galaxies.  
When observed, fields were imaged twice within the same night, once in $g$ band and once in $I$ band, with $\simeq$1 hr between the pair of observations.  This approach was chosen to provide $g-I$ color information while rejecting asteroids (by requiring two co-spatial detections) and minimizing filter changes.  The observation pairs were repeated with a cadence varying between one night and two weeks (depending on weather and the aforementioned scheduling considerations; see Fig. \ref{fig:cadence}).
The schedule was manually adjusted each night to maintain similar cadence for all fields targeted by the experiment (fields with longer elapsed time since a previous observation were given the higher priority).  The particular set of fields scheduled for a given night was chosen to minimize slew and airmass, but also anticipated the schedule for the following nights.
The effective areal coverage of the survey is shown in Fig. \ref{fig:controltime}.

\begin{figure}
\includegraphics[width=\columnwidth]{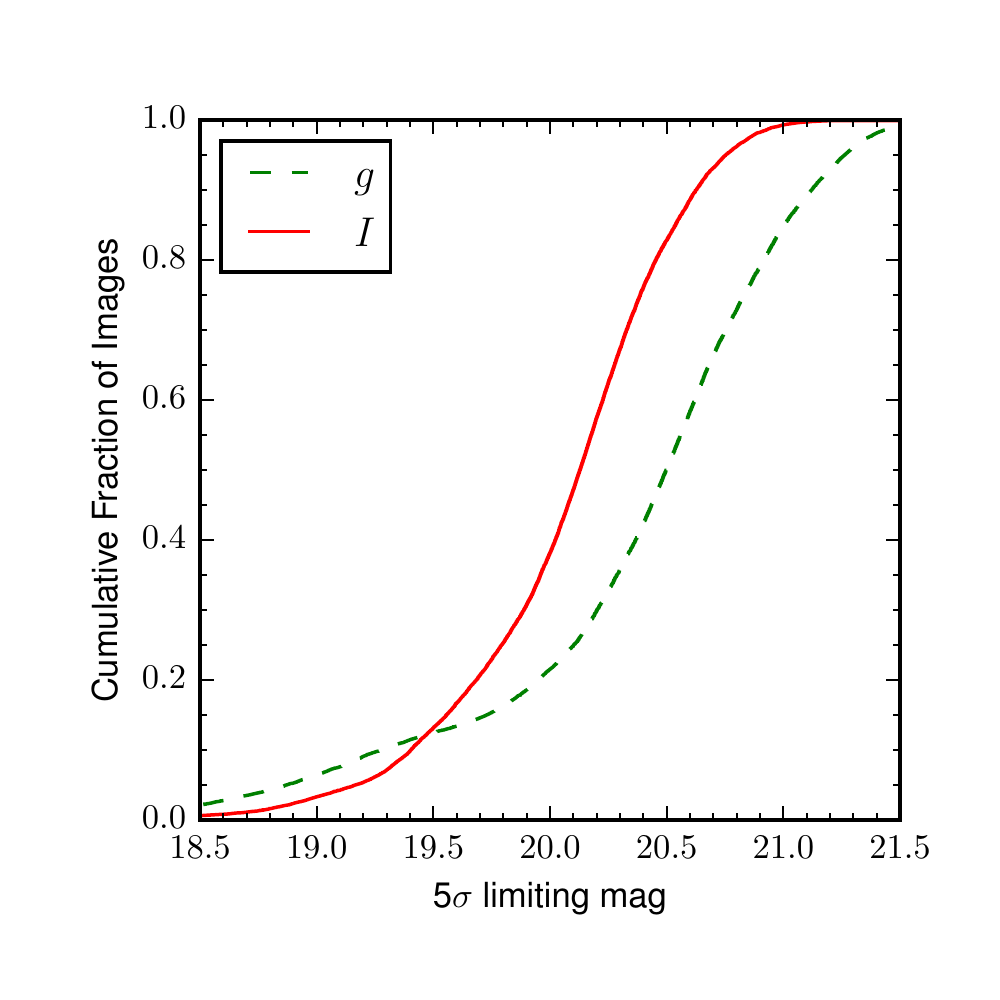}
\caption{5$\sigma$ point source limiting depth of $g$ (green dashed line) and $I$ (red solid line) band imaging during the experiment. \label{fig:lim_depth}}
\end{figure}

\begin{figure}
\includegraphics[width=\columnwidth]{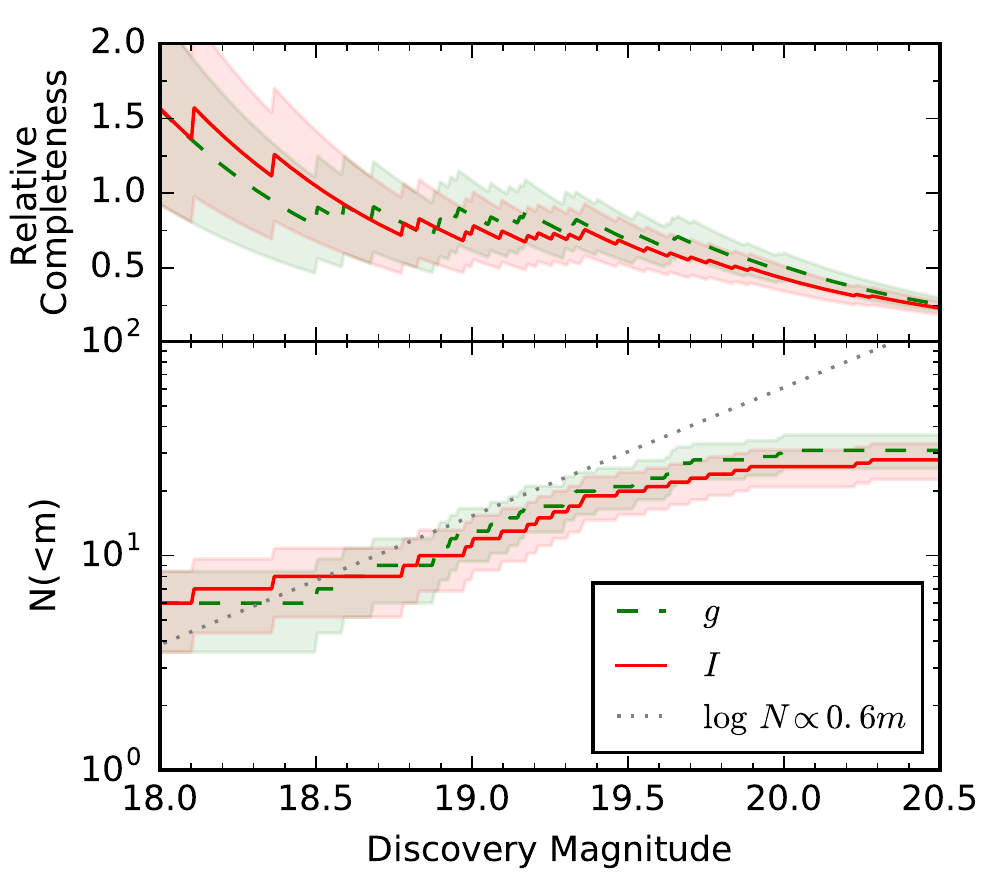}
\caption{\emph{Bottom panel:} Cumulative distributions of the  $g$ (green dashed line) and $I$ (red solid line) band magnitudes at discovery for the transients.  The shaded regions give the 1$\sigma$ Poisson ``uncertainties'' corresponding to N sources.  For comparison, the gray dotted line shows the expected cumulative counts (normalized to counts for $<19.3$ mag) if the  survey completeness was independent of magnitude (the volume probed by a magnitude-limited survey is proportional to $10^{0.6m}$).  \emph{Top panel:} relative survey completeness (observed/expected counts).  Survey completeness begins to degrade significantly for $\gtrsim 19.5$ mag.\label{fig:mags}}
\end{figure}

The survey data were processed independently with the IPAC \citep{Masci17} and NERSC realtime image subtraction \citep{Cao16} pipelines.  
Machine learning software, known as \texttt{Real-Bogus}, was used to separate real transients from image subtraction artifacts \citep{Bloom12,Brink13,Rebbapragada15}.  Sources with two detections within 24 hr (essentially requiring detections in both $g$ and $I$ bands\footnote{One transient, iPTF16jdl, was only detected in $g$-band because the $I$ band image subtraction failed for this field but was still selected because there were detections on successive nights in $g$ band observations separated by just under 24 hr.}) predicted by \texttt{Real-Bogus} to likely be real transients were then examined by human scanners.  Sources with positions consistent with a known AGN/QSO or stellar source (based on SDSS classification and past history of variability within the PTF dataset) were discarded.  The median blank-sky 5$\sigma$ point source limiting magnitudes are 20.5 and 20.1 mag in $g$ and $I$ bands respectively (Fig. \ref{fig:lim_depth}).  
However, the cumulative distributions of the discovery magnitudes of the transients (Fig. \ref{fig:mags}) shows that the survey completeness to transient sources begins to degrade significantly for magnitudes fainter than $\sim$19.5. 
The volume probed by a magnitude-limited survey is proportional to $10^{0.6m}$, where $m$ is the limiting magnitude.  Thus for a non-cosmological extragalactic survey (where the target density is independent of distance and K corrections are negligible) the cumulative number of detected sources  should be proportional to $10^{0.6m}$ and a cumulative distribution starting to fall below $\propto 10^{0.6m}$ indicates diminishing completeness.

\section{Spectroscopic Classification and Follow-up}
We spectroscopically classified all transients with $|g-I| > 0.5$ (at peak brightness)\footnote{In addition to targeting red transients, we also target blue transients to search for super-luminous SNe.}, that occurred in a known nearby ($<$200 Mpc) galaxy, or were brighter than 19.5 mag in either filter.  Spectroscopic observations were performed with the Spectral Energy Distribution Machine \citep[SEDM;][]{BlagorodnovaSEDM} on the Palomar 60-inch robotic telescope (P60), the Double Beam Spectrograph \citep[DBSP;][]{Oke82} on the 200-inch Hale telescope at Palomar Observatory (P200), the Low Resolution Imaging Spectrograph \citep[LRIS;][]{Oke95} on the 10-m Keck I telescope, DEep Imaging Multi-Object Spectrograph (DEIMOS) \citep{Faber03} on the Keck II telescope, the DeVeny Spectrograph \citep{Bida14} on the Discovery Channel Telescope (DCT), the Double Imaging Spectrometer on the Astrophysics Research Consortium 3.5-m Telescope, the Andalucia Faint Object Spectrograph and Camera (ALFOSC) on the Nordic Optical Telescope (NOT), the Gemini Multi-Object Spectrograph on the Gemini North Telescope, and the Device Optimized for the LOw RESolution (DOLORES) on the Telescopio Nazionale Galileo (TNG).  Spectral classification was done with \texttt{SNID} \citep{Blondin07} and \texttt{Superfit} v.3.5 \citep{Howell05}.

For select transients we supplemented the survey light curves from the P48 with $gri$ photometry from the Rainbow camera (RC)  on the SEDM and the GRB Cam \citep{Cenko06} on the P60.   
The RC photometry was extracted using the automatic reference-subtraction photometry pipeline,  \texttt{FPipe}
\citep{Fremling16}, which utilizes reference images of the host galaxy from the Sloan Digital Sky Survey
to subtract the background light.  
Additional photometry was obtained for iPTF17lf using the Reionization And Transients Infra-Red \citep[RATIR;][]{Fox12}, camera
mounted on the robotic 1.5-m Harold Johnson telescope of the Observatorio Astron\'omico Nacional.
For the RATIR data, the SN and the host galaxy were fitted simultaneously by modeling
the two using the point-spread function of the image
and a S\'ersic profile \citep{Sersic63}.
We obtained two epochs of \textit{BVgri} images for iPTF17bkj from Las Cumbres Observatory's 1-meter telescope in Texas \citep{Brown13}. Using \texttt{lcogtsnpipe} \citep{Valenti16}, we measured PSF photometry and calibrated it to the APASS 9 \citep{Henden16} and SDSS 13 \citep{SDSSCollaboration16} catalogs for \textit{BV} and \textit{gri}, respectively.
Supplemental photometry for iPTF17be was obtained with LRIS and with the NIRC2 near-IR camera on the Keck II telescope.

\section{Results}
We identified 36 extragalactic transients during the experiment (see Table \ref{tab:catalog}) and secured spectroscopic classifications for 34 of these: 23 SNe Ia, 5 SNe II, 2 SNe Ic, 1 SN Ib, 1 SN Ibn/IIn, 1 cataclysmic variable (CV), and 1 likely luminous blue variable (LBV) outburst.  
The sample is spectroscopically complete for the 31 of these transients brighter than 19.5 mag in either filter (20 SN Ia, 5 SN II, 2 SN Ic, 1 SN Ib, 1 SN Ibn/IIn, 1 CV, and 1 LBV outburst).  The results of the spectroscopic follow-up are summarized in Table \ref{tab:specsum}.  The sample is also spectroscopically complete for transients with $|g-I| > 0.5$ mag or coincident with a known nearby ($<$200 Mpc) galaxy, but these criteria did not select any transients not already included in the magnitude-limited sample.  We also note that the host galaxies of 4 transients (not including the CV) with redshifts corresponding to $<200$ Mpc did not have redshifts listed in the NASA Extragalactic Database\footnote{https://ned.ipac.caltech.edu}.  Though no LRNe were detected the color distribution of the detected transients confirms that a red color selection will filter out most other transients\footnote{However, care most be taken in interpreting the color distribution of the discovered transients given the requirement that sources be detected in both $g$ and $I$ bands.}.
All classification spectra have been made publicly available on the Transient Name Server\footnote{https://wis-tns.weizmann.ac.il} and the Open Supernova Catalog\footnote{https://sne.space/}.

In the following subsections we highlight three of these transients that are of particular interest: iPTF17lf (a highly reddened SN Ia), iPTF17bkj (a rare transititional SN Ibn/IIn), and iPTF17be (likely an LBV outburst).

\begin{table*}
\begin{threeparttable}
\setlength{\tabcolsep}{2pt}
\caption{Catalog of transients \label{tab:catalog}}
\begin{tabular}{lccccccccc}
\toprule

{Name} & {RA} & {Dec} & {Classification} & {Spectrum} & {Telescope/} & {Redshift} & {Peak} & {MJD} & {$g-I$} \\
       &      &       &                  &  {MJD} & {Instrument} &       & {$g$ Mag}  & {at peak} & {at peak} \\
\midrule
iPTF16iyy$^1$ & 06:56:34.60 &  +46:53:38.7 & SN II$^2$ & - & - & 0.03 & $18.42\pm0.12$ & 57763.8 & $0.19\pm0.22$ \\
iPTF16jdl$^2$ & 12:27:39.98 &  +40:36:52.7 & SN Ia$^4$ & - & - & 0.02 & $18.58\pm0.07$ & 57785.1 & - \\
iPTF16jhb$^1$ & 04:00:25.51 &  +20:19:36.2 & SN Ia & 57809.4 & P200/DBSP & 0.03 & $17.40\pm0.06$ & 57786.7 & $-0.47\pm0.14$ \\
iPTF17be$^3$ & 08:13:13.38 &  +45:59.28.9 & LBV$^7$ & 57765.3 & APO/DIS & 0.00 & $18.09\pm0.12$ & 57771.9 & $0.43\pm0.16$ \\
iPTF17jv & 08:15:04.50 &  +57:10:50.0 & SN Ia & 57779.5 & Keck I/LRIS & 0.08 & $19.47\pm0.08$ & 57770.9 & $<-0.44$ \\
iPTF17le$^1$ & 11:10:27.22 &  +28:41:42.2 & SN Ia$^8$ & - & - & 0.03 & $16.92\pm0.07$ & 57772.0 & $0.20\pm0.18$ \\
iPTF17lf$^4$ & 03:12:33.59 &  +39:19:14.2 & SN Ia$^4$ & 57774.2 & Gemini-N/GMOS & 0.01 & $19.42\pm0.08$ & 57786.6 & $1.77\pm0.13$ \\
iPTF17lj$^5$ & 07:39:01.54 &  +50:46:05.4 & SN Ia & 57780.3 & P60/SEDM & 0.03 & $17.24\pm0.05$ & 57782.9 & $-0.66\pm0.12$ \\
iPTF17ln$^2$ & 11:38:33.71 &  +25:23:49.7 & SN Ia$^8$ & - & - & 0.03 & $16.45\pm0.04$ & 57781.0 & $-0.49\pm0.11$ \\
iPTF17nh$^2$ & 07:10:13.54 &  +27:12:10.5 & SN Ia$^4$ & - & - & 0.06 & $18.74\pm0.05$ & 57785.8 & $-0.61\pm0.20$ \\
iPTF17pq & 08:03:41.63 &  +09:50:19.5 & SN II & 57785.3 & DCT/DeVeney & 0.09 & $19.40\pm0.07$ & 57780.8 & $-0.23\pm0.13$ \\
iPTF17wj & 07:24:08.95 &  +39:51:43.6 & SN Ib & 57785.3 & P200/DBSP & 0.03 & $19.15\pm0.09$ & 57782.8 & $0.15\pm0.15$ \\
iPTF17wk & 07:08:07.02 &  +49:16:29.2 & - & - & - & - & $20.71\pm0.17$ & 57782.8 & $0.43\pm0.24$ \\
iPTF17wm & 06:30:52.76 &  +73:27:31.4 & SN Ia & 57789.3 & P200/DBSP & 0.08 & $18.96\pm0.07$ & 57782.8 & $-0.22\pm0.14$ \\
iPTF17wo & 01:50:08.81 &  +39:28:52.3 & - & - & - & - & $19.64\pm0.16$ & 57782.6 & $-0.06\pm0.21$ \\
iPTF17wp & 01:23:44.06 &  +14:46:32.1 & SN Ia & 57786.1 & P200/DBSP & 0.09 & $18.95\pm0.12$ & 57785.6 & $-0.26\pm0.19$ \\
iPTF17wr & 05:02:44.92 &  +75:01:44.6 & SN Ia & 57787.3 & P200/DBSP & 0.07 & $18.50\pm0.07$ & 57789.7 & $-0.45\pm0.13$ \\
iPTF17wu & 05:46:21.12 &  +65:05:04.3 & SN Ia & 57789.3 & P200/DBSP & 0.12 & $19.96\pm0.11$ & 57771.8 & $-0.02\pm0.23$ \\
iPTF17xe$^5$ & 12:34:17.41 &  +44:18:08.9 & SN Ia & 57785.5 & P200/DBSP & 0.14 & $19.24\pm0.10$ & 57787.0 & $-0.19\pm0.17$ \\
iPTF17xf & 10:45:39.12 &  +54:07:46.3 & SN Ic & 57785.3 & P200/DBSP & 0.06 & $19.99\pm0.08$ & 57782.9 & $0.64\pm0.16$ \\
iPTF17yw & 11:47:35.37 &  +52:33:01.1 & SN Ia & 57785.5 & P200/DBSP & 0.15 & $20.33\pm0.17$ & 57783.0 & $-0.06\pm0.28$ \\
iPTF17zb$^1$ & 08:02:49.31 &  +56:37:38.8 & SN Ia & 57807.4 & P200/DBSP & 0.07 & $18.46\pm0.09$ & 57785.9 & $-0.46\pm0.16$ \\
iPTF17zp & 04:26:45.85 &  +74:04:11.5 & SN Ia & 57786.2 & P200/DBSP & 0.08 & $19.10\pm0.08$ & 57789.6 & $0.29\pm0.14$ \\
iPTF17aak & 08:30:54.62 &  +11:40:58.6 & SN Ia & 57786.4 & P200/DBSP & 0.13 & $19.18\pm0.07$ & 57789.8 & $-0.51\pm0.14$ \\
iPTF17aal & 07:43:02.52 &  +11:38:20.4 & SN II & 57786.4 & P200/DBSP & 0.07 & $18.77\pm0.07$ & 57789.8 & $-0.36\pm0.11$ \\
iPTF17aam & 07:38:40.01 &  +09:22:03.7 & SN Ia & 57793.7 & DCT/DeVeney & 0.15 & $19.87\pm0.06$ & 57785.8 & $-0.36\pm0.16$ \\
iPTF17aan & 07:39:26.00 &  +13:55:09.7 & CV & 57786.3 & P200/DBSP & 0.00 & $19.17\pm0.08$ & 57785.8 & $-0.71\pm0.16$ \\
iPTF17atf & 07:58:33.37 &  +11:55:22.0 & SN Ia & 57800.3 & P60/SEDM & 0.04 & $18.88\pm0.06$ & 57800.8 & $0.27\pm0.10$ \\
iPTF17atg & 07:39:32.53 &  +18:24:46.9 & SN Ia & 57800.3 & P60/SEDM & 0.08 & $18.98\pm0.08$ & 57800.7 & $-0.46\pm0.18$ \\
iPTF17aua & 08:59:38.73 &  +56:52:44.2 & SN Ia & 57800.5 & P60/SEDM & 0.03 & $17.47\pm0.06$ & 57799.9 & $-0.37\pm0.13$ \\
iPTF17aub & 06:40:24.70 &  +64:33:02.7 & SN II & 57800.2 & P60/SEDM & 0.01 & $17.47\pm0.05$ & 57799.8 & $-0.02\pm0.14$ \\
iPTF17avb & 08:01:15.98 &  +11:01:58.6 & SN II & 57801.2 & P60/SEDM & 0.10 & $19.53\pm0.06$ & 57800.8 & $0.44\pm0.10$ \\
iPTF17avc & 08:36:47.39 &  +45:11:35.7 & SN Ia & 57828.3 & P200/DBSP & 0.08 & $18.92\pm0.12$ & 57800.9 & $-0.34\pm0.21$ \\
iPTF17axg$^1$ & 12:36:30.75 &  +51:38:40.0 & SN Ic-BL & 57809.4 & P200/DBSP & 0.06 & $18.51\pm0.11$ & 57800.0 & $0.33\pm0.16$ \\
iPTF17bgb & 13:23:06.73 &  +52:05:57.2 & SN Ia & 57809.4 & P200/DBSP & 0.10 & $18.89\pm0.06$ & 57809.0 & $-0.67\pm0.13$ \\
iPTF17bkj$^6$ & 11:23:27.43 &  +53:40:45.4 & SN Ibn/IIn & 57809.4 & P200/DBSP & 0.03 & $19.12\pm0.08$ & 57809.0 & $0.14\pm0.14$ \\
\bottomrule
\end{tabular}
\begin{tablenotes}
\small
\item Names are cross-identified on the Transient Name Server
(TNS) as iPTFxx $=$ AT (or SN) 20xx.
Transient discoveries and classifications first reported to TNS by other groups are indicated by footnotes in the ``Name'' and ``Classification" columns respectively as follows: 
$^1$ATLAS, $^2$ASAS-SN, $^3$LOSS, $^4$TNTS, $^5$Pan-STARRS, $^6$PIKA, $^7$LCOGT-KP, $^8$PESSTO.
    \end{tablenotes}
  \end{threeparttable}
\end{table*}


\begin{table*}
\begin{threeparttable}
\setlength{\tabcolsep}{8pt}
\caption{Summary of transients \label{tab:specsum}}
\begin{tabular}{lccccccccc}
\toprule

{Classification} & {All} & {$g<19.5$} & {$g$ or $I<19.5$} & {$g-I>0.5$} & {$g-I<-0.5$} & {Known Galaxy} & {$z<0.045$} \\
       			& at peak	&    at peak  		  &     at peak  			  &     at peak        &   	peak	   &  	$<200$ Mpc      &     	     \\
\midrule
SN Ia	&	23	&	20	&	20	&	1	&	4	&	5	&	8	\\
SN II	&	5	&	4	&	5	&	0	&	0	&	1	&	2	\\
SN Ib	&	1	&	1	&	1	&	0	&	0	&	1	&	1	\\
SN Ic	&	2	&	1	&	2	&	1	&	0	&	0	&	0	\\
SN Ibn/IIn	&	1	&	1	&	1	&	0	&	0	&	1	&	1	\\
LBV		&	1	&	1	&	1	&	0	&	0	&	1	&	1	\\
CV		&	1	&	1	&	1	&	0	&	1	&	0	&	1	\\
Unclassified	&	2	&	0	&	0	&	0	&	0	&	0	&	-	\\
\hline
Total	&	36	&	29	&	31	&	2	&	5	&	9	&	14	\\
\bottomrule
\end{tabular}
  \end{threeparttable}
\end{table*}

\subsection{iPTF17lf: A Highly Reddened SN Ia}
\label{sec:17lf}
\begin{figure*}
\begin{center}
\includegraphics[width=0.9\textwidth, angle=0]{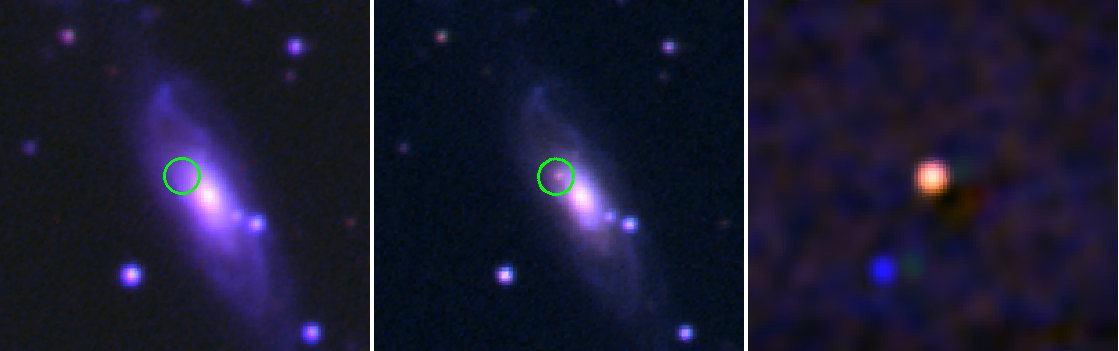}
\end{center}
\caption{Color ($gI$) reference (left), discovery (center), and smoothed difference (right) images for iPTF17lf.  The transient location is given by the green circle, which has a 5$\arcsec$ radius.
\label{fig:17lf_stamps}}
\end{figure*}

\begin{figure*}
\includegraphics[width=2\columnwidth]{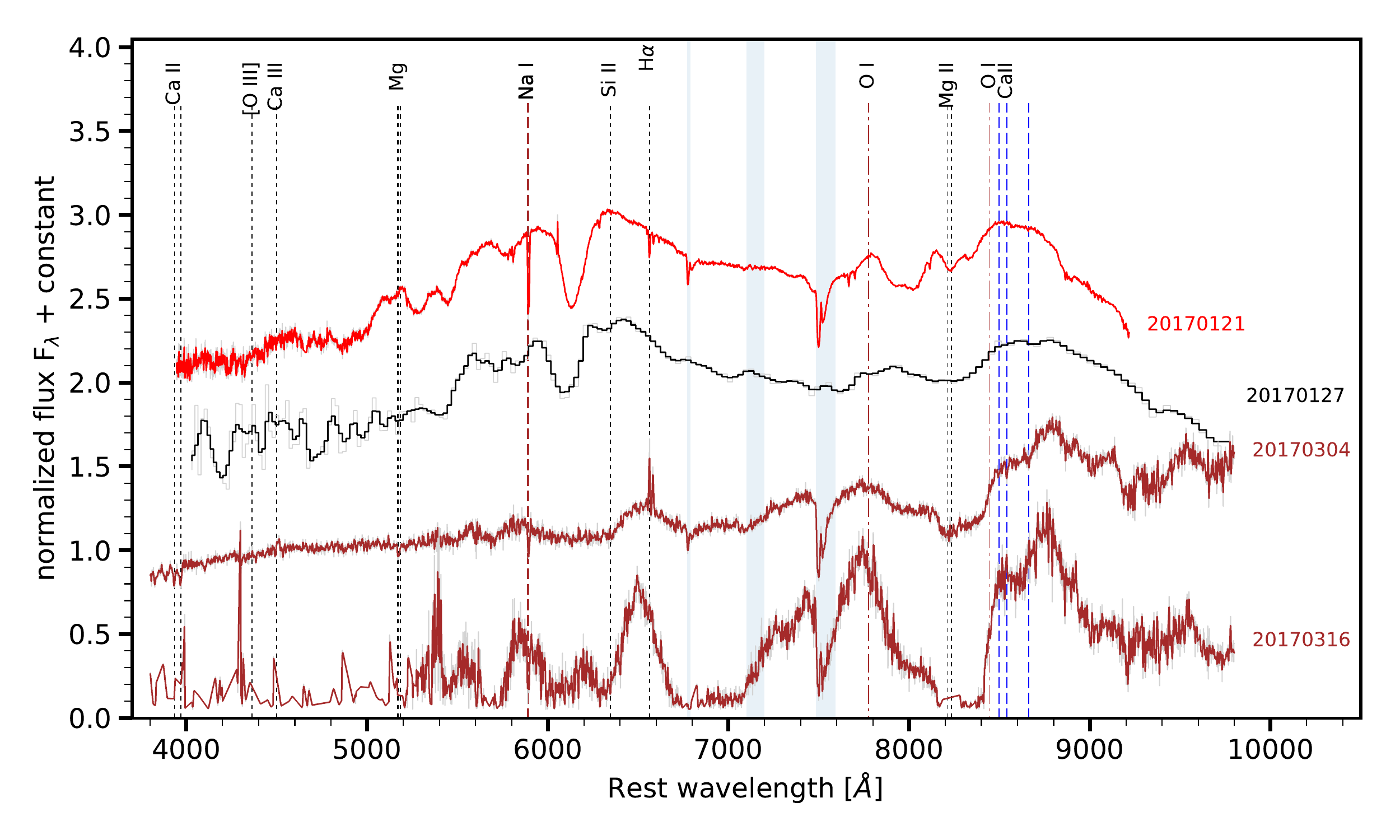}
\caption{Spectral sequence of iPTF17lf. The spectra has been smoothed using a box size of 2 pixels. A correction for the Galactic extinction has been applied. We identify the most prominent lines in the spectrum. The spectra of iPTF17lf are color coded as follows: P200+DBSP - brown, Gemini+GMOS - red, P60+SEDM - black. Shaded areas show the region with the strongest telluric absorption lines. \label{fig:17lf_spectra}}
\end{figure*}

iPTF17lf (SN 2017lf) is a highly reddened SN Ia in NGC 1233 at $\alpha$=3h 12$'$33.59$''$ $\delta$=+39$^{\circ}$19$'$14.2$''$ (J2000), discovered by the survey on 2017 Jan 18 \citep[and independently by the Tsinghua-NAOC Transient Survey on Jan 23;][]{Rui17}.  Figure \ref{fig:17lf_stamps} shows the location of the transient.  The spectral evolution of iPTF17lf is shown in Figure \ref{fig:17lf_spectra}.

Taking advantage of the standard candle property of SNe~Ia, we fit a model for the spectral energy distribution of iPTF17lf based
on the normal and unreddened SN\,2011fe as described in \citet{Amanullah15}, together with the
extinction law from \cite{Cardelli89}, to the data of iPTF17lf presented here.
The free lightcurve parameters are the time of maximum, $t_0$, and the peak brightness, $m_B$, in the rest-frame
$B$ band.  We also fit the lightcurve stretch, $s$, of the temporal evolution of the lightcurve model, which is used in
combination with $m_B$ when SNe~Ia are used to measure cosmological distances
\citep[see e.g. ][for a review]{Goobar11}.
For the host galaxy extinction we fit the color excess,  $E(B-V)$, and the total-to-selective extinction,
$R_V=A_V/E(B-V)$, in the $V$ band.  Note that we only fit for the extinction law in the host, while we adopt the
Galactic average value of $R_V=3.1$ and $E(B-V) = 0.134$ mag, based on the \citet{Schlafly11}
recalibration of \citet{Schlegel98}, for the foreground reddening in the Milky Way.

\begin{figure}
\centering
\includegraphics[width=\columnwidth]{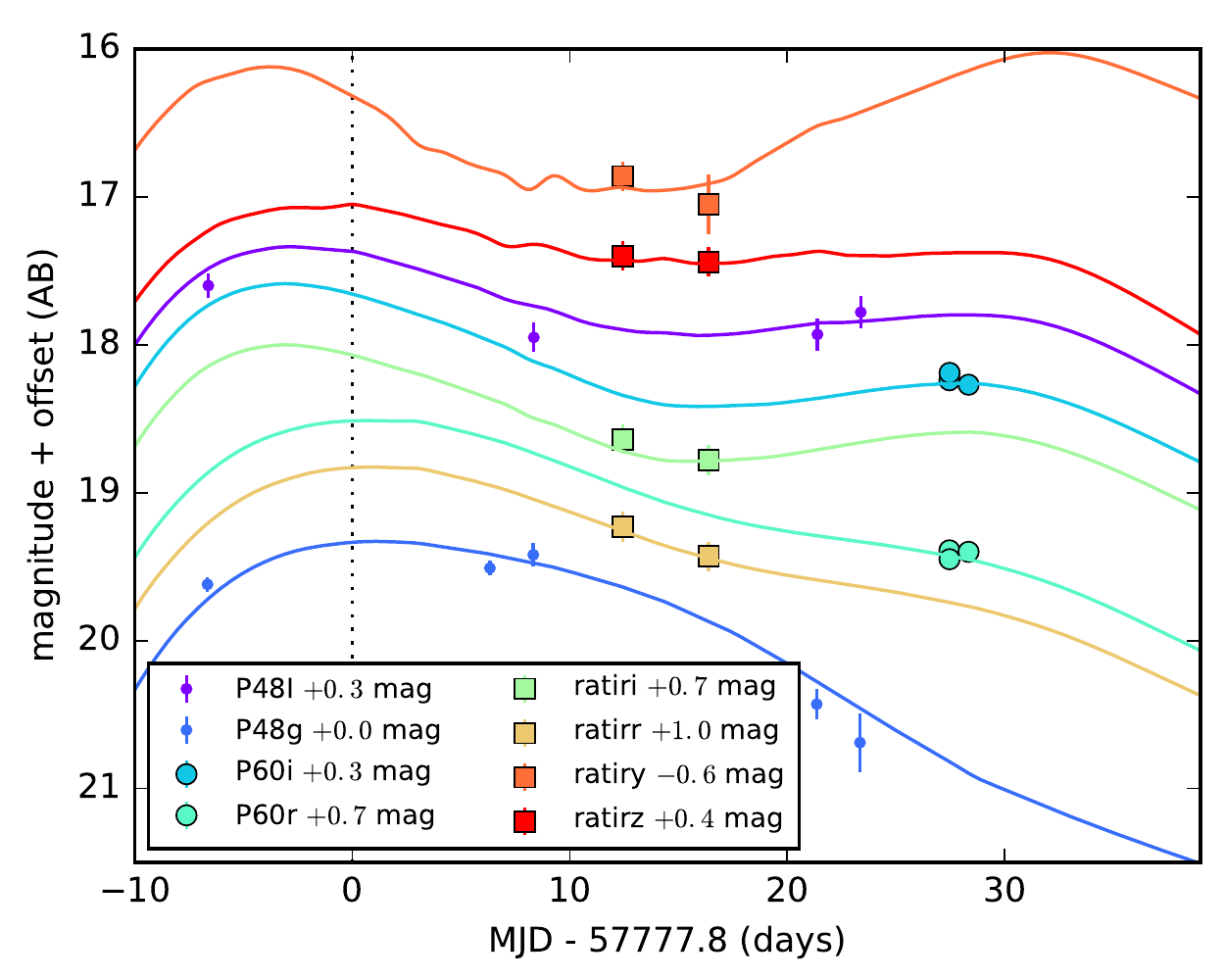}
\caption{%
        The observed lightcurve of iPTF17lf together with the fitted model (solid lines) based on SN~2011fe as described
        in \S\ref{sec:17lf}.
        \label{fig:lc17lf}}
\end{figure}

The best fit is shown together with the data in Fig.~\ref{fig:lc17lf} where the fitted parameter values were
$t_0=57777.8\pm0.5$~days, $m_B=20.34\pm0.27$~mag and $s=1.01\pm0.06$.
For the host extinction we obtain
$E(B-V)=1.9\pm0.1$~mag and $R_V=1.9\pm0.3$,
where intrinsic SN~Ia uncertainties have been taken into account.

iPTF17lf is another SN~Ia with high reddening and low $R_V$ \citep[see e.g.][]{Burns11,Amanullah15}.  Although lower $R_V$ values for highly reddened SNe is consistent with expectations for SNe discovered by magnitude limited surveys if extinction laws are drawn from an $R_V$ distribution, an alternative interpretation is that such SNe~Ia arise from progenitors surrounded by circumstellar dust \citep{Wang05,Goobar08}.  
This alternative interpretation has been further motivated by the fact that the average color correction of SNe~Ia in cosmological samples is significantly lower than what is observed in the Milky Way \citep[see e.g.][]{Betoule14}. 
If the observed $R_V$ is lower due to scattering of the SN light by circumstellar dust, the SN should appear to have to time-dependent reddening \citep[e.g.][]{Amanullah11,Brown15}.

However, this is not seen for iPTF17lf.  Although its lightcurve is sparsely sampled, it is consistent with a
constant reddening law over $\sim$35~days.  The same was seen for the highly reddened SN~2014J that was
consistent with a low value of $R_V=1.4$ \citep{Amanullah14}.  
Not only did constant reddening imply a dust distance of $\gtrsim30$ pc \citep{Bulla18}, polarimetry measurements of SN~2014J also showed the extinction to mainly be associated with the interstellar medium in the
host galaxy \citep{Patat15}.  Following the methodology of \citep{Bulla18}, the constant reddening of iPTF17lf implies a dust distance of $>$14 pc.

\subsection{iPTF17bkj (SN~2017hw): A Transitional SN Ibn/IIn}

\begin{figure*}
\begin{center}
\includegraphics[width=0.9\textwidth, angle=0]{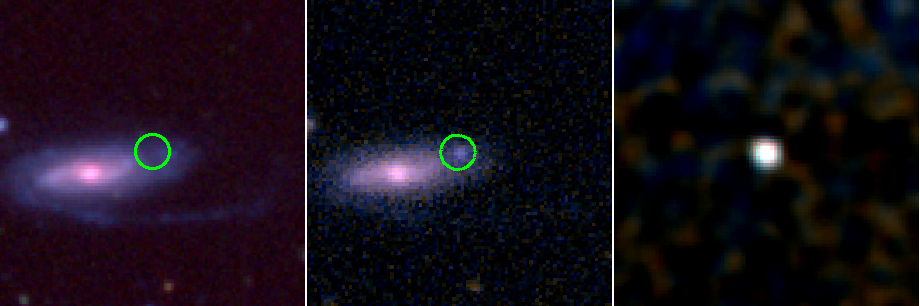}
\end{center}
\caption{Color ($grI$) reference (left), discovery ($gI$; center), and smoothed difference (right) images for iPTF17bkj.  The transient location is given by the green circle, which has a 5$\arcsec$ radius.
\label{fig:17bkj_stamps}}
\end{figure*}

iPTF17bkj is a transitional SN Ibn/IIn in UGC 6400 at $\alpha$=11$^{\mathrm{h}}$ 23$'$27.43$''$, $\delta$=+53$^{\circ}$40$'$45.4$''$ (J2000), discovered by the survey on 2017 Feb 24. An independent discovery was also reported by the Slovenian Comet and Asteroid Search Program (PIKA) on 2017 Feb 25. UGC 6400 is a star-forming galaxy at a luminosity distance of 117\,Mpc (distance modulus 35.35 mag using NED). Figure \ref{fig:17bkj_stamps} shows the location of the SN in the host galaxy.

SNe IIn are commonly accepted as supernovae interacting with their hydrogen-rich circumstellar medium (CSM). The narrow emission lines in their spectra are produced by the recombination in the surrounding shell, which is moving at lower speed than the ejecta. SNe Ibn are far less common. Similar to SNe Ib/c, they lack hydrogen features in their spectra, but they do show narrow He lines, which are believed to originate in a helium-rich CSM \citep{Pastorello07}. Few supernovae have been detected sharing the characteristics of both SNe IIn and SNe Ibn. These are the so-called transitional SNe IIn/Ibn, presenting both narrow He and H lines, with low expansion velocities ($\sim$2000\,\kms). In the literature, we could identify three objects with similar spectral characteristics: SN~2005la \citep{Pastorello2008MNRAS}, SN~2010al and SN~2011hw \citep{Pastorello15,Smith12}, which we include as our comparison sample.

iPTF17bkj was detected close to its peak light, at $M_g=-16.3\pm0.2$\,mag and $M_I=-16.4\pm0.1$\,mag (correcting for Galactic extinction of $A_V$=0.043 mag) --- $\sim$3 magnitudes fainter than the average peak magnitude for the SN Ibn population \citep{Hosseinzadeh17}. 
Dust extinction cannot be estimated from the Na I 5890\AA, 5896\AA  doublet because of He II emission, but the color of iPTF17bkj ($B-V\sim0.3$ mag at $+$15-20 d) is slightly bluer than the sample of unobscured Ibn (at similar phase) discussed in \citet{Pastorello15}, suggesting that the fainter peak magnitude of iPTF17bkj is not due to host or circumstellar dust extinction.

\begin{figure}
\includegraphics[width=\columnwidth]{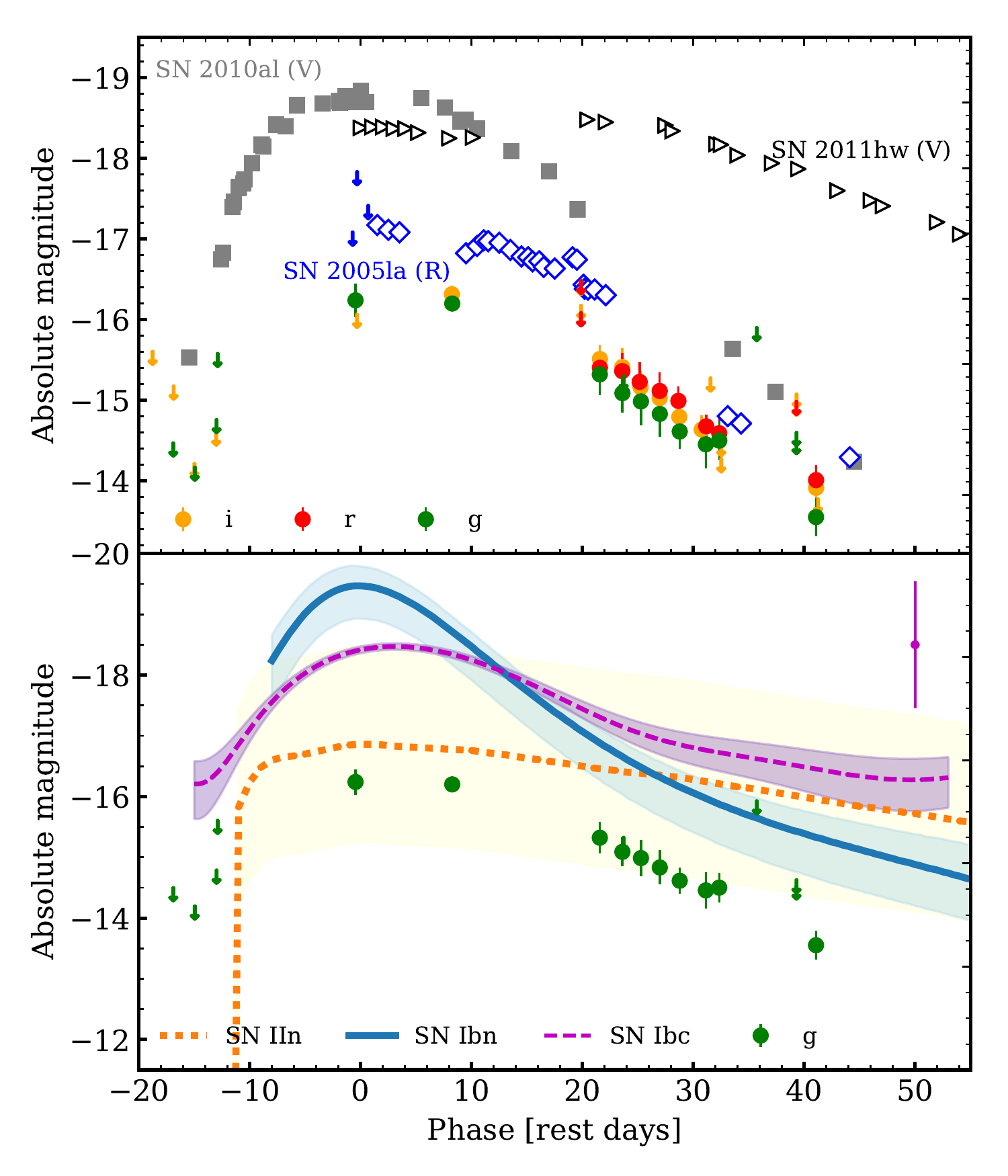}
\caption{Top panel: Colored circles show the $g$ (green), $r$ (red), $B$ (blue), $V$ (tan), and $I$ band (orange) lightcurve of iPTF17bkj. Downward arrows show the upper limits in each band. The photometry has been corrected for Galactic extinction. For comparison we also show the lightcurves of the transitional SN Ibn SN 2010al (filled gray squares) and 2011hw (open black triangles; \citealt{Pastorello15}; $V$ band) and SN 2005la (open blue diamonds; \citealt{Pastorello2008MNRAS}; $R$ band).  
Bottom panel: A comparison of the $g$-band light of iPTF17bkj to the average lightcurve for SN Ibn from \citet{Hosseinzadeh17} (blue line) with 1$\sigma$ uncertainties, the template lightcurve for  SN~Ibc in $g$ band from \citet{Taddia15} purple dashed line), and the template lightcurve for SN IIn from \citet{Li11} (orange dotted line).  The SN~Ibc template was normalized to peak magnitude (and scaled to the mean peak magnitude) and only shows the shape of the lightcurve.  The spread in the peak SN~Ibc peak magnitudes is given by the purple errorbar. 
iPTF17bkj is subluminous as compared to the sample of SN Ibn and transitional SN Ibn/IIn. Its decay timescale matches that of SN Ibc, is longer than that of SNe Ibn, and is shorter than the typical SN IIn.}
\label{fig:17bkjlightcurve}
\end{figure}

Comparing the iPTF17bkj lightcurve (see Fig. \ref{fig:17bkjlightcurve}) with an average SN Ibn template lightcurve shows not only that the peak differs, but also that the decay timescales are also longer, which implies a larger circumstellar mass \citep{Karamehmetoglu17}.
The light curve most closely resembles a subluminous SN Ibc. Individual lightcurves for the other three transitional objects seems to point to a large diversity in the distribution of the H and He CSM around these stars. One possibility, discussed by \citet{Smith12}, is that the progenitors of such explosions are LBVs transitioning to Wolf-Rayet stars.

\begin{figure*}
\includegraphics[width=2\columnwidth]{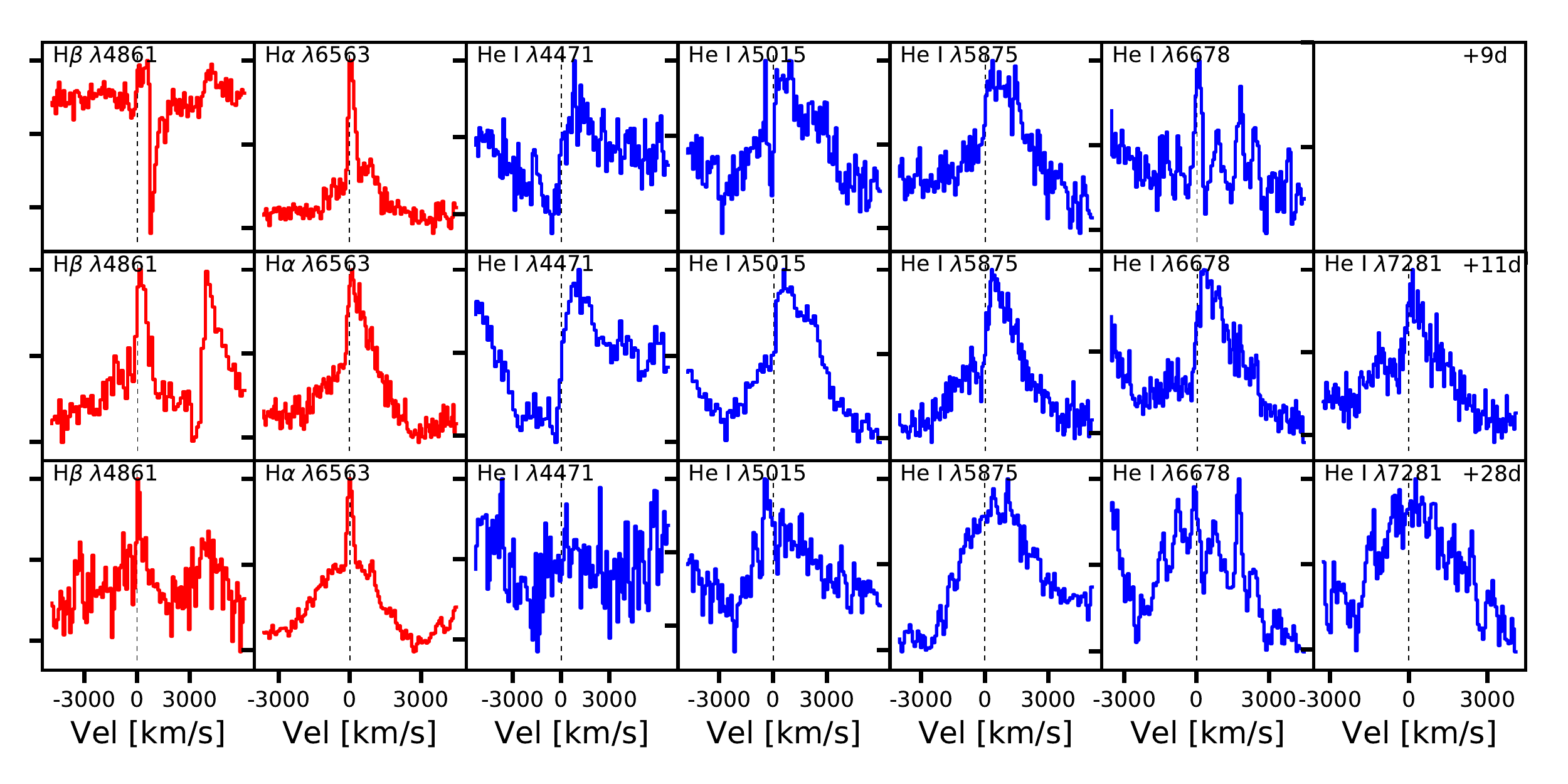}
\caption{Evolution of the H$\beta$, H$\alpha$ (red) and He I lines (blue) profiles for iPTF17bkj. Only the spectra with the best signal-to-noise have been selected. \label{fig:alt17bkj_H_profile}}
\end{figure*}

\begin{figure*}
\includegraphics[width=2\columnwidth]{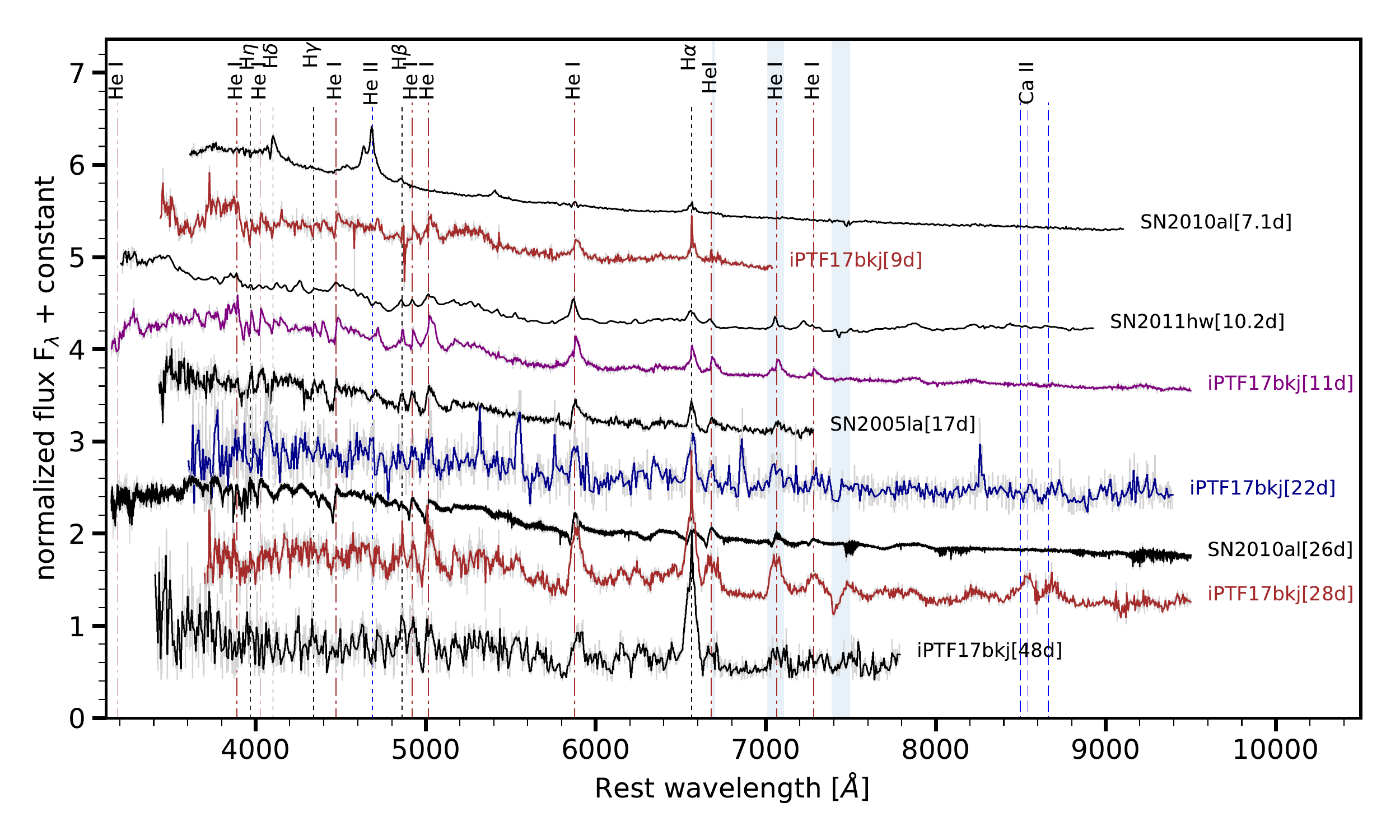}
\caption{Spectroscopic evolution of iPTF17bkj. A correction for the Galactic extinction has been applied. The dates since discovery are shown in square brackets. We also note some of the strongest emission lines. The spectra of 17bkj are color coded as follows: P200+DBSP - brown, Keck I+LRIS - purple and NOT+ALFOSC - dark blue. Although telluric lines have been removed, some residuals may be still present. Shaded areas show the region with the strongest telluric lines. As a comparison, we include spectra of other transitional SN IIn/Ibn: SN~2005la \citep{Pastorello2008MNRAS}, SN~2010al and SN~2011hw \citep{Pastorello15}. \label{fig:alt17bkj_spectra}}
\end{figure*}

The spectroscopic evolution of iPTF17bkj  is similar to that of other transitional SN IIn/Ibn objects. Figure \ref{fig:alt17bkj_spectra} shows that, at early stages ($<$15 days), the objects show narrow H$\alpha$, H$\beta$ and He I lines with Lorentzian profiles caused by electron scattering in the outer layers of the surrounding material. In the case of iPTF17bkj, we notice a narrow absorption component in both H and He I lines, with a blueshifted velocity of $\sim$400\,\kms. At later times ($>$15 days), the lines develop into a wider, P-Cygni profile, with a minimum around 2000\,\kms (see Fig. \ref{fig:alt17bkj_H_profile}). This emission is associated with the expanding photosphere of the SN, rather than diffusion of photons through the dense CSM. Already at +22\,d, we also notice the appearance of broad Ca II lines, which were not detected in the spectrum of SN~2010al taken at a similar epoch. The log of spectroscopic observations is available in Table \ref{tab:speclog}.

\begin{table*}
\begin{minipage}{1.\linewidth}
\begin{small}
\caption{Log of spectroscopic observations of iPTF17lf, iPTF17bkj, and iPTF17be.}
\centering
\begin{tabular}{cccccccc}
\hline
Object & MJD & Slit$^a$ & Telescope+Instrument & Grism$/$Grating & Dispersion & Exposure   \\ 
   & (d) &  (arcsec)  		&  	  &	& (\AA/pix)		&	(s)	     \\ \hline \hline
17lf & 57774.2 & -1 & Gemini+GMOS & $-$ & 0.9 &  900 \\ 
17lf & 57780.2 & 2.5$^a$ & P60+SEDM & IFU  &  & 2700 \\ 
17lf & 57816.1 & 1.5 & P200+DBSP & 600/4000 & 1.5 &  300 \\ 
17lf & 57828.1 & 1.0 & P200+DBSP & 600/4000 & 1.5 &  1200 \\ \hline
17bkj & 57809.4 & 1.5 & P200+DBSP & 600/4000 & 1.5  & 900 \\ 
17bkj & 57811.4 & 1.0 & Keck I+LRIS & 400/3400+400/8500 & 2.0 &  420 \\ 
17bkj & 57822.0 & 1.5 & NOT+ALFOSC & gr4 & 3.5 &  4600 \\
17bkj & 57828.3 & 1.5 & P200+DBSP & 600/4000 & 1.5 &1800 \\
17bkj & 57848.0 & 1.0 & TNG/DOLORES & LR-B &  & 3000 \\
\hline
17be & 57761.4 & 2.4$^a$ & P60+SEDM & IFU &   & 2100 \\ 
17be & 57765.3 & 1.5 & APO+DIS & B400 & 1.8  & 900 \\ 
17be & 57779.5 & 1.5 & Keck I+LRIS & 600/4000 & 1.3 & 300 \\ 
17be & 57895.2 & 1.0 & Keck II+DEIMOS & 600/6000 & 1.2 & 2100 \\
\hline
\end{tabular}
\begin{tablenotes}
    \item[\textdagger]$^a$ For the IFU, the extraction radius (in arcsec) is indicated.
    \end{tablenotes}
\label{tab:speclog}
\end{small}
\end{minipage}
\end{table*}

\begin{table*}
\begin{threeparttable}
\setlength{\tabcolsep}{2pt}
\caption{iPTF17be Progenitor Photometry \label{tab:17be_phot}}
\begin{tabular}{ccccc}
\toprule
{Filter} & {Magnitude} & $\nu L_{\nu}/L_{\odot}$ & {Telescope} & {Date} \\
\midrule
$g$ &	$>$18.2 & $<1.6\times10^6$ & P48 & 2011-04-04 -- 2014-01-03\\
$r$ &   $>$18.2 & $<1.3\times10^6$ & P48 & 2012-11-21 -- 2012-11-25\\
$I$ &	$>$18.0 & $<1.2\times10^6$ & P48 & 2016-12-14 -- 2016-12-20\\
F160W &	$>$19.76 & $<3.1\times10^4$ & NIC3/\emph{HST} & 2002-10-14 \\ 
3.6~$\mu\mathrm{m}$ & $>$17.38 & $<3.1\times10^4$ & IRAC/\emph{SST} & 2005-04-16$^a$ \\
4.5~$\mu\mathrm{m}$ & $>$17.26 & $<1.8\times10^4$ & IRAC/\emph{SST} & 2005-04-16$^a$ \\
5.8~$\mu\mathrm{m}$ & $>$15.37 & $<5.1\times10^4$ & IRAC/\emph{SST} & 2005-04-16$^a$ \\
8.0~$\mu\mathrm{m}$ & $>$13.50 & $<1.2\times10^5$ & IRAC/\emph{SST} & 2005-07-01$^a$ \\
\bottomrule
\end{tabular}
\begin{tablenotes}
\small
\item 
$^a$ -- Mean Date of multi-epoch coadded image
\end{tablenotes}
\end{threeparttable}
\end{table*}


\begin{figure*}
\begin{center}
\includegraphics[width=0.9\textwidth, angle=0]{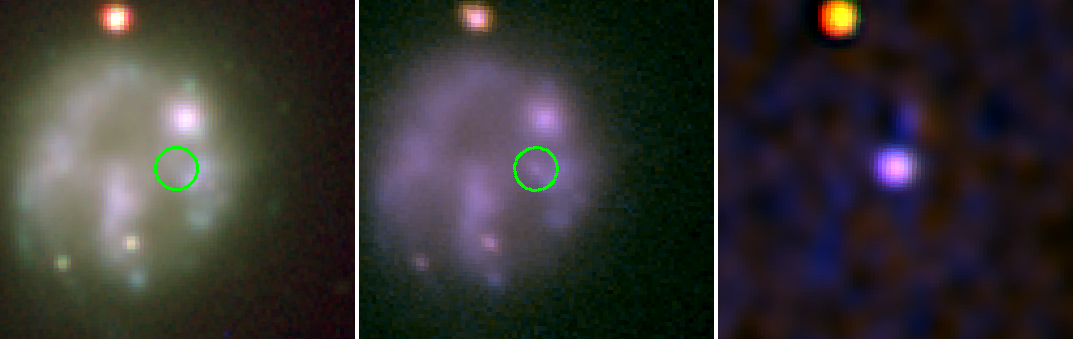}
\end{center}
\caption{Color ($grI$) reference (left), discovery ($gI$; center), and smoothed difference (right) images for iPTF17be.  The transient location is given by the green circle, which has a 5$\arcsec$ radius.
\label{fig:17be_stamps}}
\end{figure*}

\begin{figure*}
\includegraphics[width=2\columnwidth]{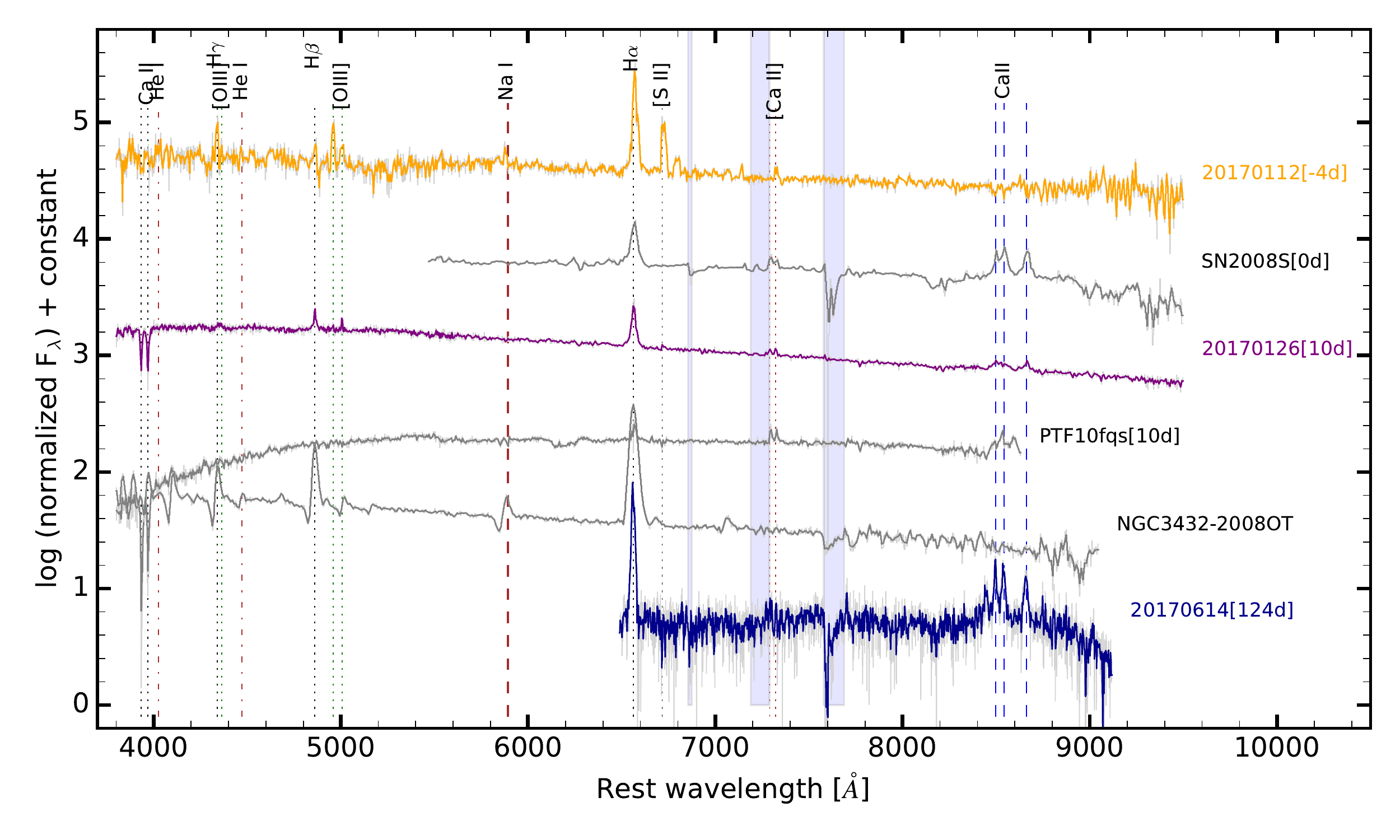}
\caption{Spectroscopic evolution of iPTF17be. A correction for the Galactic extinction has been applied. The dates since discovery are shown in square brackets. We also note some of the strongest emission lines. The spectra of iPTF17be are color coded as follows: APO+DIS - yellow, Keck I+LRIS - purple and Keck II+DEIMOS - dark blue. Although telluric lines have been removed, some residuals may be still present. Shaded areas show the region with the strongest absorption lines. As a comparison, we include spectra of other likely LBV outbursts and SN impostors: SN~2008S \citep{Botticella09}, PTF10fqs \citep{Kasliwal2011ApJ}, and the LBV in NGC 3432 during its outburst in 2008 \citep[NGC3432-2008OT;][]{Pastorello2010MNRAS}.
\label{fig:17be_spectra}}
\end{figure*}

\begin{figure}
\includegraphics[width=\columnwidth]{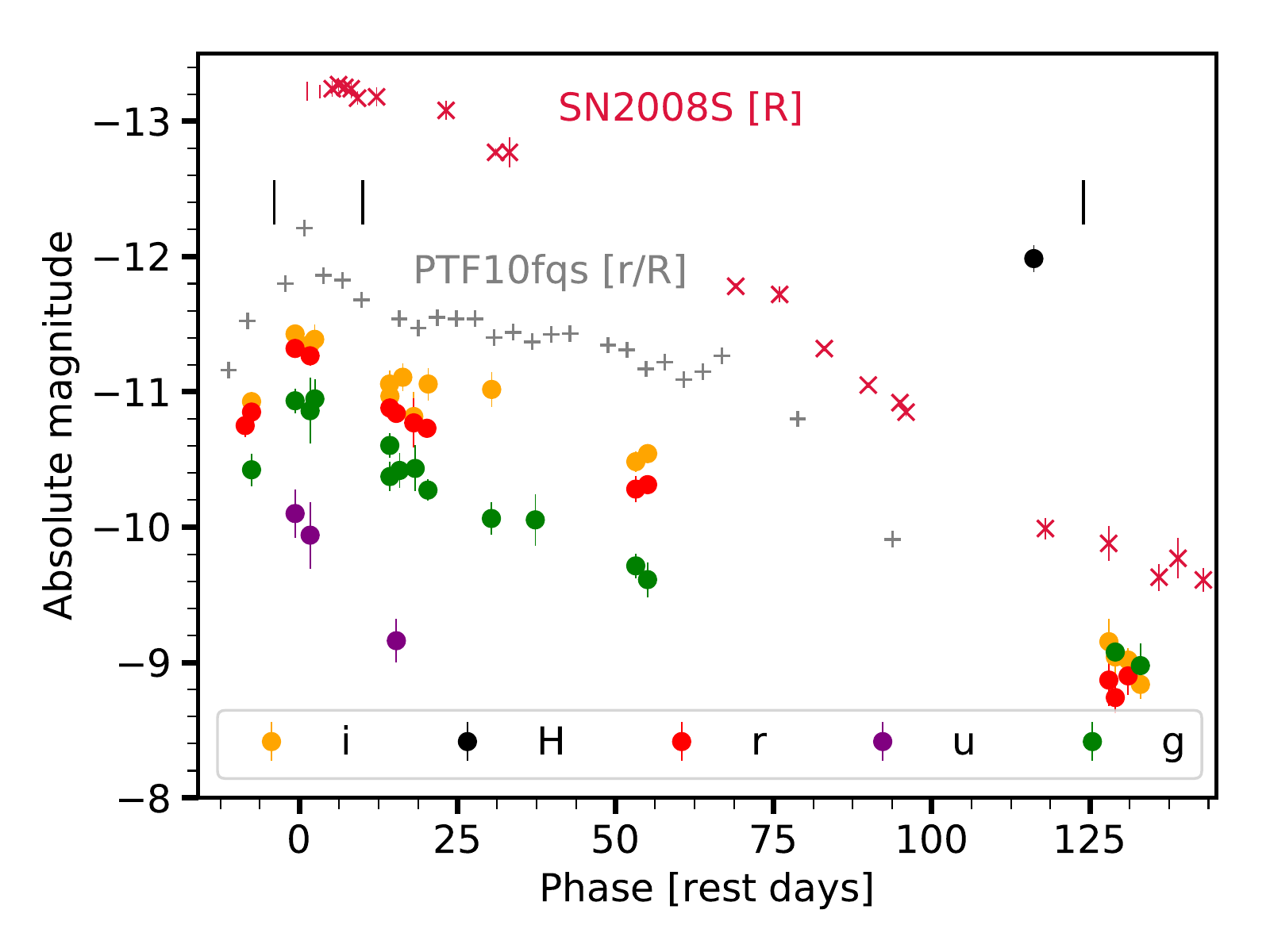}
\caption{Colored dots show the $u$ (purple), $g$ (green), $r$ (red), $i$-band (orange) lightcurve of iPTF17be. The vertical marks on top indicate the time the spectra were taken. The measurements have been corrected for foreground Galactic extinction. As a comparison, grey \texttt{+} symbols show the Mould-$R$ and SDSS $r$ band lightcurve of the ILRT PTF10fqs \citep{Kasliwal2011ApJ} and the crimson \texttt{x} symbols the $R$ band lightcurve for the LBV/ electron capture SN~2008S. \citep{Smith09}. \label{fig:17be_lc}}
\end{figure}

\begin{figure}
\includegraphics[width=\columnwidth]{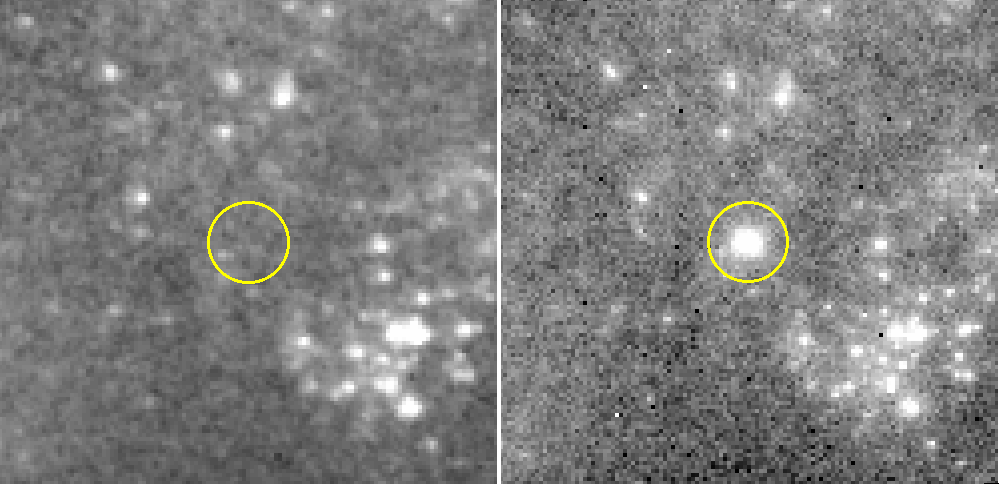}
\caption{Pre-explosion \emph{HST}/NICMOS F160W ($H$ band) imaging of iPTF17be (left) compared to late-time ($\sim$113 days post-peak brightness) Keck/NIRC2 $H$ band imaging (right).  The radius of the yellow circle is 1$\arcsec$.\label{fig:17be_nir}}
\end{figure}

\begin{figure}
\includegraphics[width=8.6cm, angle=0]{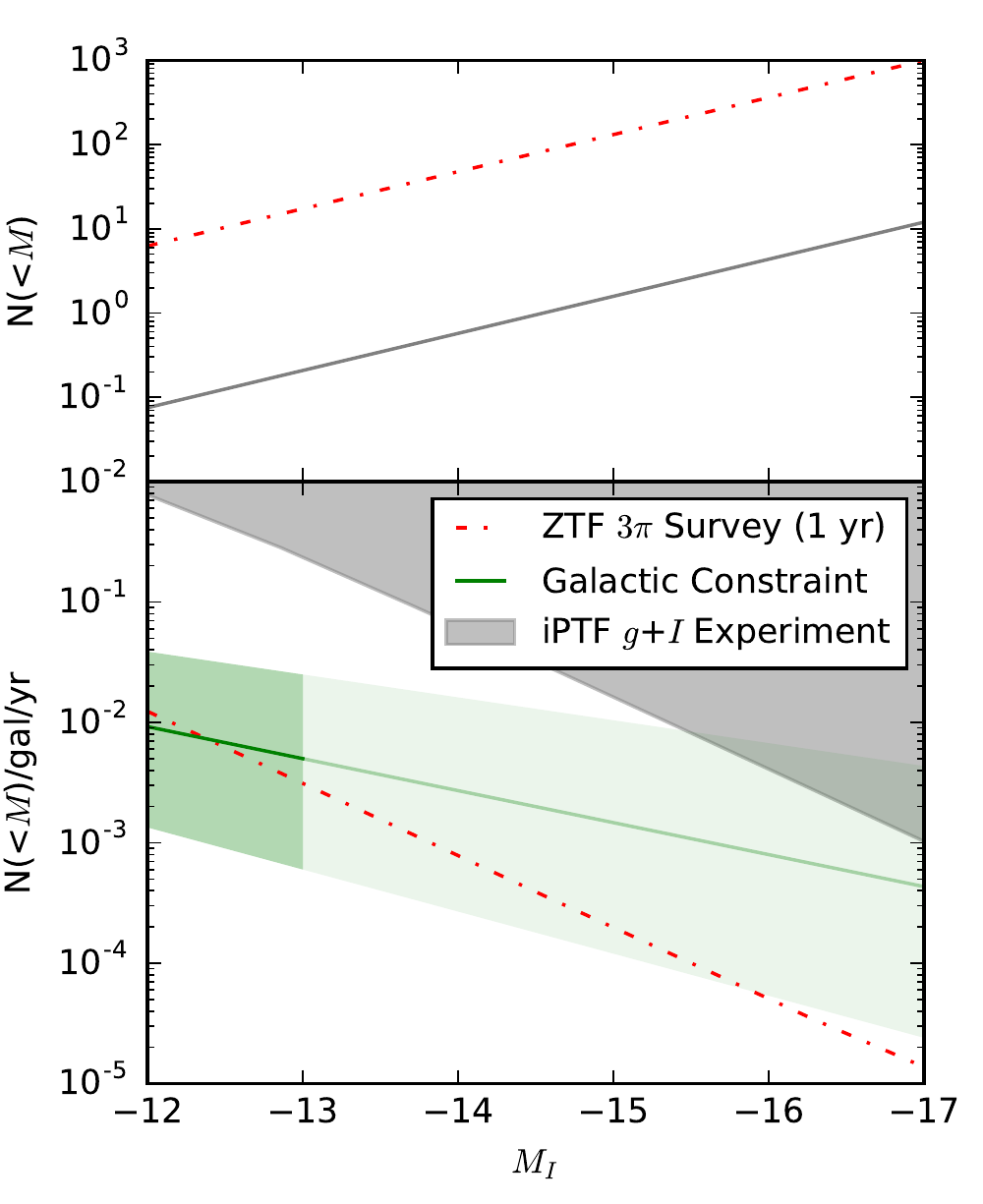}
\caption{\emph{Top:} Cumulative number of LRNe expected to be discovered by this survey (solid gray line; assuming a limiting magnitude of $m_{I}=20$ and an effective coverage of 2170 square degree months) and in the first year of the public ZTF $3\pi$ survey (dot-dashed red line; assuming 15,000 square degree years) as a function of luminosity based on the Galactic constraint from \citet{Kochanek14d}. 
\emph{Bottom:} Luminosity function of LRNe, with the rates inferred from Galactic mergers by \citet{Kochanek14d} given by the green line and 90\% confidence interval given by the green shading, and an extrapolation to higher luminosities given by the lighter green line and shading.  The 90\% confidence upper limit from this survey is shown by the gray shaded region.  The red dot-dashed line shows the corresponding constraint that would be produced by first year of the public ZTF $3\pi$ survey if it discovers zero LRNe.  The rate constraint presented in this paper only comes into tension with the Galactic constraint at very high luminosities, but the upcoming ZTF survey will be able to set more interesting constraints. \label{fig:ratelimit}}
\end{figure}

\subsection{iPTF17be (AT2017be): A Likely LBV Outburst}
iPTF17be was discovered independently by LOSS (\citealt{Stephens17}) and our survey on 07 Jan 2017 at $\alpha$=08$^{\mathrm{h}}$ 13$'$13.38$''$, $\delta$=+45$^{\circ}$59$'$28.9$''$ (J2000). The discovery image is shown in Fig. \ref{fig:17be_stamps}.  The host galaxy, NGC 2537, is located at a distance of 6.19$\pm$0.47\,Mpc (distance modulus of $\mu$= 28.96$\pm$0.16\,mag) as obtained from the redshift value reported by \citet{Garrido04}. In our analysis, we adopt a Galactic extinction of $E(B-V)=0.047$ mag from \citet{Schlafly11}.

The transient was discovered a week before reaching maximum light, as shown by the lightcurve in Fig. \ref{fig:17be_lc}. At peak, its absolute magnitude was $M_I=-11.43\pm 0.05$ and $M_g=-10.95\pm0.14$ mag, well fainter than supernovae. 
Our measurements show that the lightcurve faded quickly in the $u$ band ($\sim$1 mag in 14 days). However, slower decline was registered in redder filters ($\sim$1 mag in 28, 52, and 54 days for $g$, $r$, and $I$ bands, respectively).  The color at peak was  relatively red, $g-I\sim0.4$\,mag, as compared to the $g-I$ evolution of most supernovae (see Fig. \ref{fig:gmini}). Overall, the evolution of the $g-I$ color through $\sim100$ days post-explosion is much more modest than typically seen in LRNe or SN~2008S-like transients, which show $g-I$ color larger than 1\,mag.

The location of iPTF17bkj was observed by iPTF in previous seasons and there is also archival \emph{Spitzer} and \emph{HST}/NICMOS pre-outburst imaging of the host (see Fig. \ref{fig:17be_nir}).  
We do not detect a point-source at the location of iPTF17bkj in any of these data.
The limiting magnitudes are provided in Table \ref{tab:17be_phot} (the upper limits from the iPTF data are set from the detection of an extended source).  
Unfortunately these limits are not deep enough to set meaningful constraints on the progenitor.

The spectroscopic evolution of iPTF17be is shown in Fig. \ref{fig:17be_spectra}, along with some comparison spectra.
The first classification spectrum for this event was obtained on 08 Jan 2017 \citep{Hosseinzadeh17atel}. Despite its low signal-to-noise ratio, the classification report describes H$\alpha$ emission with a FWHM$\sim$950\,\kms in a spectrum taken 10 Jan 2017, and uses this and the low ($\sim$$-$11 mag) peak absolute magnitude to conclude that the event was likely an LBV outburst.  
Our spectrum taken near peak light (12 Jan 2017) also shows strong H$\alpha$ emission with a width consistent with that reported by \citet{Hosseinzadeh17atel}, the forbidden Ca II doublet in emission, and the Ca II triplet $\sim$8500\AA in absorption.  By the spectrum from 26 Jan 2017 (+10\, days) the triplet shifted to being in emission with broad line profiles.  
In the final spectrum taken 14 June 2017, the forbidden Ca II doublet is no longer visible and the H$\alpha$ emission is double-peaked.
The narrow H$\alpha$ and Ca triplet emission are common not only to LBV outbursts, but also to stellar mergers \citep{Rushton05,Williams15} and SN~2008S-like transients \citep{Bond09,Smith09}.  However, in contrast to iPTF17be, in well-studied stellar mergers the H$\alpha$ emission disappears and strong TiO absorption bands emerge within the first 100 days.  Since the photometric evolution of iPTF17be more closely resembles that of LBV outbursts than of SN 2008S-like events
we agree with the initial assessment of \citet{Hosseinzadeh17atel} that this event is likely an LBV outburst.  

\section{Discussion}
We now consider the implications of the results of this experiment for future surveys.
We can make constraints on the rates of various transient types using the effective survey coverage for the typical duration of each transient type (see Fig. \ref{fig:controltime}) and a limiting magnitude of 19.5 (the threshold for spectroscopic completeness).
We first check that our inferred rates of core-collapse and Type Ia SNe are consistent with values reported in the literature.  Scaling for a core-collapse SN rate of $10^{-4}~\mathrm{Mpc}^{\mathbf{-3}}\mathrm{yr}^{-1}$ \citep{Horiuchi11} and a characteristic absolute magnitude $<-16.5$ \citep{Li11b} for 30 days (corresponding to an effective survey coverage of 3731 sq-deg months) we would expect to discover 6.3 core-collapse SNe $<19.5$ mag, consistent (within Poisson statistics) with the nine we did find.  Adopting a SN Ia rate of 34\% of the core-collapse SN rate \citep{Li11b} and adopting a characteristic absolute magnitude for SNe Ia of $<-18$ for 10 days (corresponding to an effective survey coverage of 2170 sq-deg months), we would expect 19.8 SN Ia brighter than 19.5 mag, again consistent with the 20 observed.

We can now make a constraint on the rate of LRNe based on the lack of discoveries during the survey.
The number, $N$, expected in a flux limited survey, is 
\begin{equation}
N\propto \int \frac{dN}{dL} L^{3/2} dL
\end{equation}
since the volume probed goes as $L^{3/2}$.  For our survey the expected number depends on the luminosity function for high luminosities ($M<-14$), but constraints on the luminosity function have only been estimated using the more common, lower-luminosity events discovered within the Galaxy, where \citet{Kochanek14d} found the Galactic rate of LRNe to be $dL/dN\propto L^{-1.4\pm0.3}$ with $\sim$0.1--1 yr$^{-1}$ brighter than $M_{I}=-4$ mag and that the peak luminosities are roughly 2000-4000 times brighter than the main-sequence progenitor luminosities.

Clearly this luminosity function must steepen for very high luminosities (the expected rate in a flux limited survey is only finite for luminosity functions steeper than $dN/dL \propto L^{-2.5})$, otherwise transient surveys would routinely detect distant, luminous LRNe.  Instead, only a handful have been discovered, with only three beyond the local group.  
Although stellar mergers are expected to eject mass primarily at velocities no more than a few times the escape velocity (making them distinguishable from typical SNe regardless of their luminosity), it has been conjectured that some stellar mergers may become dust-obscured prior to their peak luminosity \citep{Metzger17} and would be discovered as infrared transients \citep{Kasliwal17} rather than by an optical survey as LRNe.  If the peak luminosities of LRNe are a few thousand times the progenitor main-sequence luminosity the most luminous LRNe would be roughly -16 to -18 mag. 

Even the non-detection of any LRNe during just this two-month experiment is in tension with an extrapolation of the stellar merger luminosity function inferred from Galactic events to very high luminosities ($M_{I}\lesssim -16$ mag), but it is still consistent with the peak luminosities of LRNe being a few thousand times the progenitor main-sequence luminosity.  We calculate the expected number of LRNe using Eqn. 1 normalized by the Galactic rate, assuming $10^{-2}$ Milky Way-like galaxies per Mpc$^{-3}$, and scaled for a survey limiting magnitude of $m_{I}=19.5$ and an effective coverage 2170 square-degree months (the survey coverage for transients with a 10 day duration).
The top panel of Fig. \ref{fig:ratelimit} shows the cumulative number of LRNe as a function of peak luminosity that should have been found by this experiment based on this extrapolation 
and the bottom panel shows the corresponding constraint on the rate of LRNe.

Deep all-sky time-domain photometric surveys with color information, such as the public 3$\pi$, 3-d cadence $g$ and $r$ band survey of the upcoming ZTF and, later, the Large Synoptic Survey Telescope, will be capable of making much more meaningful constraints on the high-end luminosity function of LRNe.  The highest luminosity stellar merger candidate known peaked at $-$14 mag \citep{Smith16}.  The ZTF survey will have the sensitivity, within its first year, to determine whether the LRN luminosity function steepens before or after this luminosity.  

This $g+I$ band survey is a pathfinder for future transient surveys.  It was the first $I$ band transient search with PTF/iPTF and its successful implementation laid the groundwork for including an $I$ band component to ZTF, which will improve the photometric classification and selection of cool and dusty transients as well as SNe Ia.  

\section*{Acknowledgements}
We thank Mislav Balokovic, Kevin Burdge, Kishalay De, George Djorgovski, Andrew Drake, Marianne Heida, Nikita Kamraj, and Harish Vedantham for taking observations used in this paper.  We thank the anonymous referee and Chris Kochanek for helpful comments.

AAM is supported by a grant from the Large Synoptic Survey Telescope Corporation, which funds the Data Science Fellowship Program.  JS and FT gratefully acknowledge the support from the Knut and Alice Wallenberg Foundation.  RA, MB, AG, and RF acknowledge support from the Swedish Research Council and the Swedish Space Board.

The Intermediate Palomar Transient Factory project is a scientific collaboration among the California Institute of Technology, Los Alamos National Laboratory, the University of Wisconsin, Milwaukee, the Oskar Klein Center, the Weizmann Institute of Science, the TANGO Program of the University System of Taiwan, and the Kavli Institute for the Physics and Mathematics of the Universe.
Part of this research was carried out at the Jet Propulsion Laboratory, California Institute of Technology, under a contract with the National Aeronautics and Space Administration. 
Some of the data presented herein were obtained at the W.M. Keck Observatory, which is operated as a scientific partnership among the California Institute of Technology, the University of California and the National Aeronautics and Space Administration. The Observatory was made possible by the generous financial support of the W.M. Keck Foundation. 
These results made use of the Discovery Channel Telescope at Lowell Observatory. Lowell is a private, non-profit institution dedicated to astrophysical research and public appreciation of astronomy and operates the DCT in partnership with Boston University, the University of Maryland, the University of Toledo, Northern Arizona University and Yale University.  The upgrade of the DeVeny optical spectrograph has been funded by a generous grant from John and Ginger Giovale.
This work is partly based on observations made with the Nordic Optical Telescope, operated by the Nordic Optical Telescope Scientific Association at 
the Observatorio del Roque de los Muchachos, La Palma, Spain, of the Instituto de Astrofisica de Canarias. This work is partly based on observations made with DOLoRes@TNG.


\bibliography{references}
\bibliographystyle{aasjournal}
\clearpage
\end{document}